\documentclass[12pt,preprint]{aastex}

\lefthead{Skinner et al.}
\righthead{Herbig Ae Stars}

\begin{document}

\title{New Perspectives on the X-ray Emission of HD 104237 and Other \\ 
       Nearby Herbig Ae/Be Stars from {\em XMM-Newton} and {\em Chandra} }

\author{Stephen L. Skinner}
\affil{CASA, Univ. of Colorado, Boulder, CO 80309-0389 }

\author{Manuel G\"{u}del}
\affil{Paul Scherrer Institut, W\"{u}renlingen and Villigen,
CH-5235 Switzerland}

\author{Marc Audard}
\affil{Columbia Astrophysics Laboratory, Columbia Univ., 550 W. 120th St.,
       New York, NY 10027  }

\and 

\author{Kester Smith}
\affil{Max Planck Inst. f\"{u}r Radioastronomie, Auf dem H\"{u}gel 69,
       D-53121 Bonn, Germany}


%
\newcommand{\ltsimeq}{\raisebox{-0.6ex}{$\,\stackrel{\raisebox{-.2ex}%
{$\textstyle<$}}{\sim}\,$}}
\newcommand{\gtsimeq}{\raisebox{-0.6ex}{$\,\stackrel{\raisebox{-.2ex}%
{$\textstyle>$}}{\sim}\,$}}

\begin{abstract}
The origin of the X-ray emission from Herbig Ae/Be
stars is not yet known. These intermediate mass pre-main
sequence stars lie on radiative tracks and are not
expected to emit X-rays via solar-like magnetic processes, nor
are their winds powerful enough to produce X-rays by
radiative wind shocks as in more massive O-type stars.
The emission could originate in unseen low-mass companions,
or it may be intrinsic to the Herbig stars themselves
if they still have  primordial magnetic fields or can
sustain magnetic activity via a nonsolar dynamo.

We present new X-ray
observations of the nearby Herbig Ae star HD 104237 
(= DX Cha) with {\em XMM-Newton}, whose objective is to 
clarify the origin of the emission. Several X-ray
emission lines are clearly visible in the CCD spectra,
including the high-temperature Fe K$\alpha$ complex.
The emission can be accurately modeled as a  multi-temperature
thermal plasma with cool (kT $<$ 1 keV) and hot 
(kT $\gtsimeq$ 3 keV) components. The presence of a
hot component is compelling evidence that the X-rays
originate in magnetically confined plasma, either in
the Herbig star itself or in the corona of an as yet
unseen late-type companion. The X-ray temperatures
and luminosity (log L$_{X}$ = 30.5 ergs s$^{-1}$) are 
within the range expected for a T Tauri companion,
but high resolution {\em Chandra} and {\em HST}
images constrain the separation of a putative 
companion to $<$1$''$. We place these new results
into broader context by comparing the X-ray and 
bolometric luminosities of a sample of nearby
Herbig stars with those of T Tauri stars and  
classical main-sequence Be stars. We also test the
predictions of a model that attributes the X-ray
emission of Herbig stars to magnetic activity
that is sustained by a  shear-powered dynamo.
\end{abstract}




%
\newpage

\section{Introduction}
The nearest star-forming regions are dominated by populations
of young low-mass pre-main-sequence (PMS) stars, or T Tauri
stars. Hundreds of T Tauri stars have now been identified in
the Taurus-Auriga, Chamaeleon, and Rho Ophiuchus molecular clouds 
at distances of $\ltsimeq$160 pc. Their masses are 
less than $\sim$2 M$_{\odot}$ and they are the progenitors
of late-type main sequence stars such as the Sun. In contrast,
more massive PMS stars are rare in nearby star-forming regions.
These elusive objects evolve rapidly onto the main sequence and 
can remain optically obscured during much of their PMS phase.
However, a few nearby optically-bright PMS stars of intermediate
mass ($\sim$2 - 10 M$_{\odot}$) are known. They are 
generally categorized as Herbig Ae/Be stars (Herbig 1960;
Herbig \& Bell 1988),
which are thought to be more massive analogs of  classical T 
Tauri stars (CTTS). 
They are often located in or near molecular clouds in the vicinity 
of T Tauri stars and have properties similar to CTTS including
H$\alpha$ emission, infrared excesses indicative of circumstellar 
material (disks, envelopes, or both), and poorly-understood photometric 
and spectroscopic variability (Appenzeller 1994; Finkenzeller \& Mundt 
1984, hereafter FM84; Herbig 1994; Hillenbrand et al. 1992). In 
addition, mass outflows are often detected (FM84; Garrison \&
Anderson 1977; Garrison 1978;  Skinner, Brown, \& Stewart 1993).

One of the nearest known Herbig Ae stars is HD 104237 (= DX Cha),
lying in the recently discovered $\epsilon$ Cha group at a 
{\em Hipparcos} distance of 116 pc (Perryman et al. 1997; 
van den Ancker, de Winter, \& Tjin A Djie 1988, hereafter AWD98). 
This  star has attracted 
considerable observational attention because of its 
proximity, optical brightness (V = 6.6 mag), and relatively
low visual extinction A$_{V}$ $<$ 1 mag. Its properties are
summarized in Table 1.  It does not lie in a molecular
cloud but molecular gas has been detected in its vicinity
(Knee \& Prusti 1996) and age estimates range from 
$\sim$2 - 5 My (AWD98; Feigelson, Lawson, \&  Garmire 2003,
hereafter FLG03; Grady et al. 2004, hereafter G04). 
A spectral type of A4IVe was assigned in the early study of
Hu et al. (1991), but more recent observations with the
{\em Hubble Space Telescope (HST)} suggest a later spectral 
type of A7IVe (Brown et al. 1997) or A7.5Ve - A8Ve (G04).

Observations with the
{\em Infrared Space Observatory (ISO)} reveal a clear
IR excess and a strong 10 $\mu$m silicate feature
(Meeus et al. 2001). The IR excess can be modeled
in terms of a passive irradiated disk with
a small but uncertain 
outer radius of $\sim$10 - 20 AU (Dominik et al. 2003).
The {\em HST}
has traced a bipolar jet-like outflow (HH 669) with at least three
emission knots down to angular separations of $\approx$0.125$''$ from 
the star (Woodgate et al. 2002; G04). There has also been a 
report of the detection of a close companion in  {\em HST}
and infrared images at a separation of 1.365$''$ $\pm$ 0.019$''$, 
possibly a M3-4
T Tauri star (Danks et al. 2001; G04).

{\em ASCA} detected HD 104237 as a moderately bright 
X-ray source, as reported by Skinner \& Yamauchi (1996;
hereafter SY96). The X-ray detection of HD 104237 has 
recently been confirmed in two higher angular resolution
{\em Chandra} observations that were obtained as part of
a limited snaphsot survey of Herbig stars (FLG03). This
snapshot survey also detected the Herbig stars HD 100546 (B9Vne), 
HD 141569 (B9.5Ve), and HD 150193 (A1Ve). These high angular
resolution {\em Chandra} detections substantiate earlier
reports of X-ray emission from Herbig stars, including
eleven X-ray detections obtained with the {\em Einstein}
observatory (Damiani et al. 1994) and a surprisingly high 
detection rate of nearly 50\% in a sample of 30 Herbig Ae/Be 
stars observed by {\em ROSAT} (Zinnecker \& Preibisch 1994, 
hereafter ZP94; Preibisch \& Zinnecker 1996).

The apparent presence of X-ray emission from  Herbig
stars is difficult to explain in the context of
current theories of stellar X-ray emission. Herbig Be
stars and earlier Ae stars are on radiative tracks and
are not expected to have the outer convection zones
needed to sustain  internally-generated magnetic fields. 
Consequently, X-ray emission from magnetically-trapped coronal
plasma such as occurs in the Sun and other cool stars is 
not expected. However, some recent evolutionary models
such as those of Siess, Dufour, \& Forestini (2000)
suggest that later Herbig Ae stars of spectral types
A6e - A8e  could have very thin convection zones 
(Giardino et al. 2004), thus raising the interesting
possibility of incipient solar-like magnetic activity 
in later Ae stars (see also Sec. 7.3).
Finally, it is very unlikely
that the X-ray emission of Herbig Ae stars arises in 
radiative wind shocks, as may be the case for much more
massive O-type stars.  The instabilities that
are thought to produce X-ray emitting shocks in the 
supersonic winds of O stars are not predicted to occur
in the more benign winds of Ae stars, which typically 
have velocities v$_{\infty}$ $\approx$ 200 - 600 km s$^{-1}$ 
and ionized mass loss rates $\dot{M}$ $\leq$ 10$^{-7}$ 
M$_{\odot}$ yr$^{-1}$ (FM84; Skinner, Brown, \& Stewart 1993).
Furthermore, X-ray temperatures of several keV have now 
been detected in Herbig Ae stars (including HD 104237),
and such temperatures are inconsistent with the soft emission
predicted by radiative wind shock models (SY96).

A possible explanation of the X-ray emission from Herbig
stars is that it arises from faint late-type companions.
However, {\em Chandra's} arcsecond angular resolution is
now placing tight upper limits on the separation
of such putative X-ray companions from the Herbig star.
As discussed in more detail below, {\em Chandra} observations 
of HD 104237 show that the X-ray peak is offset by $<$1$''$
from the stellar optical position, thus ruling out the
faint star seen 1.365$''$ away by {\em HST} as the X-ray 
source.  Although {\em Chandra's} angular resolution is
not sufficient to exclude the presence of a cool secondary
at separations of $<$1$''$, the recent {\em HST} observations
fail to find any evidence for a T Tauri companion at
separations 0.05$''$ $<$ $r$ $\leq$ 1.0$''$ (G04). This still
does not rule out an  unseen companion at
$r$ $\leq$ 0.05$''$, but it is worth keeping in
mind that other factors could lead to intrinsic X-ray emission
from Herbig stars themselves. Of potential importance is their
extreme youth and rapid rotation. Their ages 
are usually less than a few million years (Strom et al.  1972;
Hillenbrand et al. 1992) and typical rotational velocities
are $v$sin$i$ $\approx$ 80 - 150 km s$^{-1}$ 
(Finkenzeller 1985; B\"{o}hm \& Catala 1995). Interestingly,
a much lower value $v$sin$i$ = 12 $\pm$ 2 km s$^{-1}$
has been measured for HD 104237 by Donati et al. (1997).
{\em HST} observations indicate that this low $v$sin$i$ is
probably a result of the system being viewed nearly 
pole-on at an inclination angle $i$ = 
18$^{\circ}$$^{+14}_{-11}$ (G04).
 
Because of their youth, 
primordial magnetic fields inherited from
the parent molecular cloud may still be present, and a
marginal detection of a magnetic field in HD 104237 has
been reported by Donati et al. (1997). It has also 
been suggested that the shear energy in a young
rapidly-rotating star might give rise to a nonsolar dynamo, 
allowing an internally-generated field to be sustained 
in the youngest Herbig stars that decays rapidly over 
a few million years (Tout \& Pringle 1995;  Vigneron et
al. 1990). If magnetic fields are indeed 
present, they could lead to magnetic plasma confinement
and X-ray emission.

We present here new X-ray observations of HD 104237 obtained
with {\em XMM-Newton} that were motivated by
the unanswered question of the origin of X-ray emission from
Herbig stars. The new data provide higher signal-to-noise
CCD spectra and X-ray light curves than were  previously
available. Based on the inferred plasma properties, we
argue for a magnetic  origin for the X-ray
emission. We use previously published and archived X-ray data 
for a distance-limited sample of Herbig stars to make 
comparisons with lower mass T Tauri stars and classical
Be stars. We show that if Herbig stars are the evolutionary
precursors of classical Be stars then a substantial falloff
in the X-ray luminosities of Herbig stars is expected after
they reach the main sequence, and a possible link with 
the similar behavior predicted by shear-dynamo theory is
noted.

\section{Observations}

\subsection{XMM-Newton Observations }
Table 2 summarizes the {\em XMM-Newton} observations.
Further information on the X-ray telescope is given
by Jansen et al. (2001). Our analysis is based on
CCD images, spectra, and light curves from the 
European Photon Imaging Camera (EPIC). Data were
acquired simultaneously with the  EPIC-PN camera
(Str\"{u}der et al. 2001) and two nearly identical EPIC-MOS
cameras (MOS-1 and MOS-2; Turner et al. 2001). 
The PN and MOS cameras provide
a $\approx$30$'$ diameter field-of-view and energy coverage
from $\approx$0.2 - 15 keV, moderate energy
resolution (E/$\Delta$E $\approx$ 20 - 50), and 
$\approx$6$''$ FWHM angular resolution on-axis.

Data reduction followed standard procedures using the 
{\em XMM-Newton} Science Analysis System software (SAS vers. 5.4.1).
Pipeline-processed events files generated using the most current
calibration data were filtered with {\em evselect} to select good 
event patterns. Spectra and light curves were extracted from
the filtered events lists within a circular region of radius
R$_{e}$ $\approx ~$18$''$ centered on HD 104237 (Fig. 1). This radius
corresponds to $\approx$75\% of the encircled energy at 1.5
keV. Response matrix files (RMFs) and auxiliary response files
(ARFs) tailored to the specific observational parameters were
generated using the SAS tasks $rmfgen$ and $arfgen$. The ARF
file corrects the measured source flux for energy that falls
outside of the R$_{e}$ $\approx ~$18$''$ extraction region. 

As discussed below (Sec. 3.0.1), four faint X-ray sources lie at 
separations of 4.1$''$ - 15$''$ from HD 104237.
We attempted to remove some of the contamination these 
nearby sources  by excluding
events within small circular regions centered on their positions,
which are accurately known from higher resolution {\em Chandra}
images (Fig. 2). These smaller exclusion regions are shown as 
dashed circles in Figure 1. This strategy can remove some - but
not all - of the contamination from these nearby sources.
Specifically, sources B and C (Fig. 2) lie in the wings of the {\em XMM-Newton}
point-spread function (PSF) at separations of 4.1$''$ - 5.8$''$ from 
HD 104237, and cannot be spatially resolved from the Herbig star
at {\em XMM-Newton's} spatial resolution. 

Background was
extracted from source-free regions on the detector near HD 104237.
Spectra were analyzed using XSPEC v. 11.1 and light curve analysis
was undertaken with XRONOS v. 5.18
\footnote{Further information on XSPEC and XRONOS can be found at
http://heasarc.gsfc.nasa.gov/lheasoft/xanadu. They are part of the
XANADU software package maintained by the High Energy Astrophysics
Science Archive Research Center (HEASARC) at NASA's Goddard Space Flight
Center.}. Spectra were rebinned to a minimum of 15 counts per bin for analysis 
and all  spectral models included an
absorption component based on Morrison \& McCammon (1983) cross 
sections.

\subsection{Chandra Archive Observations of HD 104237}
We have made use of two short $\approx$3 ks exposures of HD 104237
available in the {\em Chandra} public archive. Each of these exposures
provides only $\approx$400 counts for HD 104237 and they are thus of
limited use for spectral analysis. However, the images obtained with
{\em Chandra's} higher angular resolution (90\% encircled energy
radius $\approx$2$''$ at 1.5 keV) and well-calibrated boresight
provide crucial information on the position of the bright X-ray source
relative to the optical position of HD 104237 as well as
precise locations of
fainter sources within $\approx$15$''$ of HD 104237. The positions of
these fainter sources cannot be accurately measured in the lower
resolution {\em XMM} images.

The {\em Chandra} observations
were obtained with the ACIS-I detector in Faint/Timed mode on 
5 June 2001 (ObsId 2404, 2.96 ks) and 4 Feb 2002 (ObsId 3428, 
2.83 ks) with HD 104237  positioned $\approx$1.7$'$ off-axis.
Level 2 data products generated during {\em Chandra X-ray Center}
standard processing were further processed using CIAO software vers. 2.2.1
\footnote{Further information on {\em Chandra Interactive 
Analysis of Observations (CIAO)} software can be found at
http://asc.harvard.edu/ciao} in order to incorporate observation-specific
bad pixel files and apply standard aspect corrections and energy filters.
A more complete description of the {\em Chandra}
observations  can be found in FLG03, and {\em Chandra} instrumentation 
is described in the {\em Chandra Proposer's Observatory Guide (POG)}
\footnote{http://asc.harvard.edu/proposer/POG/} and in 
Weisskopf et al. (2002).

\section{Results }

\subsubsection{X-ray Images and Source Identification}

Figure 1 shows the unsmoothed full resolution {\em XMM-Newton} 
image in the immediate vicinity of HD 104237 using summed 
data from the nearly identical MOS1 and MOS2 detectors. The MOS
data provide somewhat better image quality than does PN
because of the smaller  1.1$''$ MOS pixel size, which fully
samples the telescope PSF. Figure 2
shows the same region imaged by {\em Chandra} ACIS-I on
4 Feb 2002. Both images clearly show a prominent X-ray
source near the HD 104237 optical position, and the 
{\em Chandra} image also reveals four fainter sources
within 15$''$ of HD 104237. These are labeled as 
B,C,D, and E for consistency with the notation used
in FLG03. Additional information on the optical and
X-ray properties of these four faint sources is
given in FLG03 and G04.

The first {\em Chandra} image obtained on 5 June 2001
is similar to the second image shown in Figure 2 except that
source C was not detected in the first observation.
After applying
the standard aspect correction to the first {\em Chandra}
image and correcting the {\em Hipparcos} position for
proper motion to the epoch of the {\em Chandra} observation,
the position of the 
pixel having maximum brightness
is offset by only 0.52$''$ from the {\em Hipparcos} position
of HD 104237  (Perryman et al. 1997), and by 0.56$''$
from its near-IR position in the 2MASS Point Source
Catalog.  Similarly, the offsets between the {\em Chandra}
peak positions of sources D and E and their 2MASS counterparts
identified below are 0.54$''$ and 0.55$''$,  respectively.
Our comparison of the {\em Chandra} and 2MASS images suggests
that these offsets could very likely be reduced to 
$\approx$0.2$''$ - 0.3$''$ by cross-registration.  However, an astrometric
solution was not attempted because of the small number of
X-ray sources detected in the short {\em Chandra} exposures.
Even without cross-registration, the above comparisons
indicate that the {\em Chandra} positional
accuracy relative to the {\em Hipparcos} and 2MASS frames
is already better than 0.6$''$, in agreement with results 
from {\em Chandra}  calibration studies
\footnote{http://asc.harvard.edu/cal/ASPECT/}.

Sources B and C are separated 
from HD 104237 by only 4.1$''$ - 5.8$''$ and cannot be
clearly resolved at {\em XMM's} lower spatial resolution.
Thus, our X-ray flux measurements based on {\em XMM-Newton}
spectra will contain a small contribution from these two sources.
Even so, their contribution to the total flux is expected
to be minor  since their respective {\em Chandra}
count rates were  $\leq$2\% and $\approx$7\% of the 
HD 104237 count rates. Sources D and E are offset
from  HD 104237 by $\approx$11$''$ - 15$''$ and are
visible in the {\em XMM-Newton} image as extended structure
to the southeast (Fig. 1). They were classified as
T Tauri stars by FLG03 and are visible in the near-IR
as  2MASS 12000829$-$7811395 and 2MASS 12000931$-$7811424.
A comparison of the two
{\em Chandra} observations shows that sources C and
E are variable,  but C is very faint (6 counts) and
was only detected in the second observation.

Several other X-ray sources are present in the EPIC images, as listed
in Table 3 and shown in Figure 3. The source list in Table 3
is based on a visual comparison  of the PN and MOS images
with the list of PN sources detected in the 0.5 - 4.5 keV
range as part of the standard {\em XMM-Newton} pipeline
processing. This processing uses the SAS  sliding box
detection  algorithm {\em eboxdetect} along with maximum-likelihood
task {\em emldetect} for PSF fitting.
Table 3 includes only those PN sources 
that were also confirmed to be present in one or
both MOS detectors, thus minimizing spurious detections
at the risk of omitting some faint sources. 
Candidate near-IR or optical counterparts
lying within 5$''$ of the X-ray positions were found for 9 of
the 19 sources in Table 3, two of which are late-type stars.
These identifications are based on
searches of the 2MASS all-sky Point Source Catalog and the 
SIMBAD database.

The B9Vn star $\epsilon$ Cha lies $\approx$2.2$'$ southwest of
HD 104237 and they form a common proper motion pair. Molecular
gas has been detected around $\epsilon$ Cha and it may be a 
young object (Knee \& Prusti 1996), but it lacks the emission 
lines needed to 
be classified as a Herbig Be star. We did not detect
$\epsilon$ Cha with {\em XMM-Newton}, nor was it detected
in the {\em Chandra} observations.  The upper
limit obtained from the EPIC PN image is 
log L$_{X}$ (0.5 - 7 keV) $\leq$ 27.8 ergs s$^{-1}$.
Here we have assumed a 1 keV thermal plasma spectrum
with an absorption N$_{H}$ = 1.1 $\times$ 10$^{21}$
cm$^{-2}$ corresponding to A$_{V}$ = 0.5 mag (Knee
\& Prusti 1996), and a {\em Hipparcos} distance of
112 pc. The {\em Chandra} upper limit 
log L$_{X}$ $\leq$ 27.7 ergs s$^{-1}$ (FLG03) is
slightly more stringent because of {\em Chandra's}
much lower detector background.

\subsubsection{X-ray Variability}

Figure 4 shows the background-subtracted EPIC-PN X-ray light 
curve of HD 104237 in the [0.5 - 5] keV range. We have used
this restricted energy range for light curve analysis in order
to minimize the possibility of contamination from soft and
hard background photons. 

No large-amplitude variations are visible in the PN light
curve but a slow falloff in the count rate is apparent.
This decline is also seen in the background-subtracted MOS 
light curves and in PN light curves constructed with
other energy filters including a soft filter in
the range [0.5 - 2.0] keV and a hard filter 
of [2.0 - 5.0] keV. 
The mean count rate for the PN light curve
in Figure 4 is $\mu$ = 0.25 $\pm$ 0.04 c s$^{-1}$ 
($\pm$1 $\sigma$). During the first 4 ks of the observation
the mean is $\mu$ = 0.27 $\pm$ 0.03 c s$^{-1}$ while
the last 4 ks gives  $\mu$ = 0.23 $\pm$ 0.03 c s$^{-1}$.
The probability of a constant count
rate from a $\chi^2$ analysis of the PN light curve 
binned at 200 s intervals is P$_{const}$ = 0.18
and smaller values P$_{const}$ $<$ 0.01 are obtained from
the summed MOS1 $+$ MOS2 light curve. 

As an additional check for variability, 
we have applied the Kolmogorov-Smirnov (KS) test 
(Press et al. 1992) to the unbinned PN event list, 
using events in the  [0.5 - 5] keV range from a 
contiguous good-time interval during the last  9880 s 
of the observation. For the event extraction region
shown in Figure 1, the KS test gives P$_{const}$ =
0.001. If we instead use a simpler extraction region
consisting of all events inside a small circle of radius 
R$_{e}$ = 8$''$ centered on HD 104237, then the KS
test gives  P$_{const}$ = 0.068.

The above results suggest that low-level variablity
is quite likely present in the light curve. However, the 
origin of the variability remains ambiguous because of
the  nearby faint sources that are not clearly resolved
from HD 104237 at  {\em XMM-Newton's} angular resolution.
Our analysis
shows that the apparent low-level variability is present 
using source extraction regions centered on HD 104237
with radii as small as R$_{e}$ = 6$''$ - 7$''$. Such
a small extraction region excludes most of the photons
from sources D and E, but contains photons from sources
B and C in addition to those from HD 104237. Thus, sources
D and E are not likely to be responsible for the variability.
The most probable origin would then be HD 104237 itself,
or sources B or C, or perhaps an as yet undetected 
close companion to HD 104237. We emphasize that even though the  
{\em XMM-Newton} observation has insufficient angular
resolution to rule out variability from sources B or
C, the two {\em Chandra} observations found no significant
variability in source B and showed that source C is 
very faint (FLG03).

The two   {\em Chandra} exposures  were separated by 
eight months in time and provide an additional
check for variability. Even though the exposures were short
($\approx$3 ks each), there is little or no contamination of the 
X-ray light curves by nearby sources  B,C,D, and E
because of {\em Chandra's} higher angular resolution.
The light curves show count rate fluctuations at
the 2$\sigma$ level but no statistically significant
variability was found in either {\em Chandra} observation.
However, the mean count rate during the second observation
was somewhat higher than the first. The first observation
detected 382 $\pm$ 20 counts in 2.96 ks within a circle of radius 
3.9$''$  centered on HD 104237, giving a mean rate in
the 0.5 - 7 keV range $\mu$ = 0.13 $\pm$ 
0.01 (1$\sigma$) c s$^{-1}$.
Using the same extraction region, the second observation
gave  477 $\pm$ 22 counts in 2.83 ks and
$\mu$ = 0.17 $\pm$ 0.01 c s$^{-1}$ (0.5 - 7 keV).
These count rates may slightly underestimate the true
values because of moderate photon pileup in ACIS-I.
The $\approx$30\% higher count rate during the second
observation may reflect real long-term variability, but 
the inferred change is of low statistical significance 
because of the scatter in the two light curves.

Slowly declining light curves similar to that detected 
by {\em XMM-Newton}  have been seen in other pre-main sequence stars,
such as the CTTS Haro 5-59 in Orion
(Fig. 7 of Skinner, Gagn\'{e}, \& Belzer 2003).
In some cases, the decline lasts for at least a day.
The origin of such slow variability is not yet understood
but possible explanations include the late decay phase of 
an X-ray flare that occurred prior to the observation or  
dynamical effects such as the rotation of an active region
out of the line-of-sight.  The absence of a detectable
change in  hardness ratio or temperature in the 
{\em XMM-Newton} light curve of HD 104237 suggests 
that the slow variability
was due to a change in the emission measure,
perhaps through dynamical effects. Obviously, longer
term monitoring over timescales of days to weeks is
needed to more accurately characterize the level of
the variability and further constrain its origin.

\subsubsection{XMM-Newton  Spectra}

Figure 5 shows the EPIC-PN spectrum, which provides
a higher signal-to-noise (S/N) ratio for spectral analysis
than  the MOS. Three emission line features are clearly
visible and are identified with  transitions
in  the following He-like ions,
where E$_{lab}$ is the laboratory line energy and 
T$_{max}$ (K) is the temperature at which the
line emits maximum power: 
Si XIII (E$_{lab}$ = 1.86 keV,
log T$_{max}$ = 7.0), Ca XIX (E$_{lab}$ = 
3.90 keV, log T$_{max}$  = 7.4), and 
the Fe K$\alpha$ complex which includes
Fe XXV (E$_{lab}$ = 6.67 keV,
log T$_{max}$  = 7.6). The individual
components of the He-like triplets are not
spectrally resolved by the PN or MOS detectors.
The Fe K$\alpha$  complex 
is confirmed in the lower 
S/N MOS spectra and provides unambiguous evidence
for hot plasma.

\noindent{\em Fluorescent Fe I K$\alpha$?}
In addition to the obvious line detections listed above, 
our analysis of the unbinned photon event list from the PN 
detector shows a weak feature at an energy slightly above 
6.4 keV. A Gaussian fit of this feature gives a mean energy
$\langle$E$\rangle$ = 6.46 [6.36 - 6.61; 90\% conf.] keV.
This feature is not present in the background spectrum.
If real, this feature could be a weak detection of
fluorescent emission from neutral iron (Fe I K$\alpha$) 
in cold material surrounding the star.
Such fluorescent iron lines have recently been detected by {\em Chandra}
in other hot stars such as the classical Be star $\gamma$ Cas
(Smith et al. 2004). However, we caution that: (i) the statistical
significance of this detection is low, (ii) the feature is not
seen in  the lower S/N MOS data for HD 104237, and (iii) some
blending of this feature with higher energy photons from
the Fe K$\alpha$  complex ($\approx$6.67 keV) could be present 
at the spectral resolution of the PN detector (which  is 
$\approx$0.13 keV at 6.5 keV). It is thus clear that more
sensitive observations will be needed to determine if
Fe I K$\alpha$ emission is indeed present.

\noindent {\em X-ray Temperatures and EM Distribution:} 
Acceptable fits of the spectrum were obtained using absorbed
variable-abundance multi-temperature optically thin plasma models,
as summarized in Table 4. For these fits, we used the {\em vapec} 
model as implemented in XSPEC v. 11.1. In addition, fits of
comparable  quality were obtained with the differential emission
measure (DEM) model {\em c6pvmkl}, which approximates the 
DEM using a sixth order Chebyshev polynomial.

Single-temperature (1T) {\em vapec} models were not acceptable,
but satisfactory fits were obtained using both two-temperature (2T)
and three-temperature (3T) models, with 3T models giving
slightly better fits as shown in Figure 5 (see also Table 4). The 2T
and 3T models are qualitatively similar in that they both
require cooler plasma at kT $<$ 1 keV and a hotter component 
at kT $\approx$ 3 keV. However, the 3T model requires higher
absorption and places a larger fraction of the total emission
measure (EM) in cooler plasma below 1 keV. Acceptable
fits were obtained with the 2T and 3T models using an 
Fe abundance that is consistent with solar, but an overabundance
is inferred for Ca and perhaps Ne.

The hotter component at kT $\approx$ 3 keV is anticipated
from high-temperature features such as the 
Fe K$\alpha$ line complex and was also required 
to fit previous {\em ASCA} spectra
(SY96). At lower temperatures, both the 2T and 3T models 
require a component at  kT $\approx$ 0.6 keV and the
3T model yields a slight fit improvement to the low-energy
part of the spectrum by adding a third
very soft component at kT $\approx$ 0.15 keV whose EM is
quite uncertain. This soft component is heavily absorbed 
and accounts for only $\approx$6\% of the observed (absorbed) flux
in the PN spectrum. Because of the marginal fit improvement
gained by adding this very soft component, its low
contribution to the observed flux, and calibration 
uncertainties at low energies, its physical reality is
open to question. The acceptable fits obtained by both
2T and 3T models illustrate the ambiguity that is often
present in physical interpretations of moderate resolution
CCD spectra. The discrete-temperature 2T and 3T models 
are simple approximations of what is likely to be a more
complex plasma temperature distribution.

DEM models also show a double-peaked structure with a 
cool peak below 1 keV and a hotter peak above 3 keV.
The temperatures at which these peaks occur are somewhat
sensitive to abundance assumptions but are typically 
in the range  kT $\approx$ 0.25 - 0.6 keV for the cool
component and kT $\approx$ 4 - 5 keV for the hot 
component. DEM models converge to an absorption 
N$_{H}$ $\approx$ (1.9 - 2.0) $\times$ 10$^{21}$ cm$^{-2}$,
which is consistent with the values derived from
3T models.

\noindent{\em Abundances:} 
Fits with the 2T $vapec$ model using solar abundances 
(Anders \& Grevesse 1989) were not acceptable. We thus
allowed the abundances of Ne, Mg, Si, S, Ca, and Fe to 
vary, and in doing so were able to obtain acceptable
fits (Table 4). The fit improvement obtained by using variable
abundances is significant in the 2T model but marginal 
in the 3T model. Even though improved fits were obtained
with variable abundances, it should be kept in mind that
CCD spectra from PN and MOS  provide only moderate
spectral resolution and thus cannot provide definitive
abundance estimates because of blended lines. 

Only in the case of Ca did we find a significant
departure from solar abundances, with values of
6.6 - 9.9 $\times$ solar inferred, albeit with large 
uncertainties (Table 4). This overabundance is
required to fit the line feature near 3.9 keV,
which is identified as Ca XIX. Ar XVIII 
Ly$\beta$ also emits
near this energy but we are unable to fit the 
feature by varying the Ar abundance.  Also, there
is no obvious detection of the Ar XVIII Ly$\alpha$
line near 3.3 keV, which should be several times
stronger than the line near 3.9 keV in the 
temperature range of interest here.
Thus, Ca XIX is the most likely
line identification.  Neon is the only other element
whose abundance converges to nonsolar values
at the 90\% confidence level in {\em vapec} fits.
The 2T and 3T {\em vapec} models yield a Ne abundance 
of about twice solar (Table 4).

\noindent {\em Absorption and Visual Extinction:}
Using the results
in Table 4 and the conversion from N$_{H}$ to 
A$_{V}$ of Gorenstein (1975), 
the 2T model implies a visual extinction 
A$_{V}$ = 0.35 [0.24 - 0.48] mag, while the 3T
model gives A$_{V}$ = 1.01 [0.54 - 1.40] mag,
where brackets enclose 90\% confidence intervals.
DEM models give values A$_{V}$ $\approx$ 0.9 mag
that are comparable to
the 3T model.
Optical studies based on the {\em Hipparcos} 
distance have given values in the range 
A$_{V}$ = 0.3 mag (AWD98)
to A$_{V}$ = 0.9 mag (Malfait, Bogaert, \& Waelkens
1998). Thus, the X-ray derived values are within
the range inferred from optical data.

\noindent {\em Extraction Region Comparisons:} The above
results are based on the extraction of source events
using the region shown in Figure 1. In order to
determine if the spectral analysis results are 
sensitive to the extraction region used, we extracted
a PN spectrum using a simpler region consisting of 
all events inside a small circular region of radius
R$_{e}$ = 8$''$ centered on HD 104237. Fits of this
spectrum with a 2T $vapec$ model yielded nearly identical
values of N$_{\rm H}$, kT$_{1}$, and observed flux to
those given in Table 4. The inferred value of kT$_{2}$
was $\approx$23\% less than that in Table 4, but its 
90\% confidence range overlaps that in Table 4. We thus
conclude that our spectral analysis results are robust
to changes in the source extraction region.

\subsubsection{Chandra Spectra}
We have analyzed the photon event lists for the two
{\em Chandra} observations and extracted ACIS-I
spectra as well as source-specific RMF and ARF
files using the CIAO tool {\em psextract}.
Because of the low number of counts (382 - 477 counts
per spectrum), no corrections for charge transfer inefficiency
were  applied and we made use of {\em a priori} information
on absorption and abundances from the  {\em XMM-Newton} spectra
to reduce the number of free parameters in spectral fits.

The photon energy distributions for HD 104237 are 
similar  for the two {\em Chandra} observations. Using
events in the 0.5 - 7 keV range, the mean photon energy in
the first observation was $\langle$E$\rangle$ = 1.26 $\pm$ 
0.67 (1$\sigma$) keV, while the second observation gave
$\langle$E$\rangle$ = 1.40 $\pm$ 0.83 (1$\sigma$) keV.

We have fitted the spectrum of the second {\em Chandra}
observation with a 2T {\em vapec} model, fixing the 
absorption N$_{H}$ and abundances at the values 
determined from $vapec$ model fits of the higher S/N 
{\em XMM-Newton} spectra (Table 4). The 2T $vapec$ fit
yields temperatures that are very similar to
those obtained with {\em XMM}, namely 
kT$_{1}$ = 0.76 [0.64 - 0.84] keV and kT$_{2}$ =
3.0 [2.1 - 5.6] keV, where brackets enclose 
90\% confidence intervals. The observed (absorbed)
flux of the second {\em Chandra} observation determined 
from the 2T $vapec$ model is 22\% less than that
measured by {\em XMM-Newton} (Table 4). The 3T
$vapec$ model gives somewhat better  agreement,
with the observed {\em Chandra} flux being 10\%
less than that of {\em XMM-Newton} (Table 4). 
Some (or perhaps all) of the excess flux measured
by {\em XMM-Newton} relative to {\em Chandra} no
doubt comes from the two faint nearby sources 
B and C, which cannot clearly be separated from
HD 104237  at {\em XMM-Newton's} lower spatial
resolution (Fig. 1).

\section{X-ray Data for Other Nearby Herbig Stars}
In the discussion below (Sec. 5), we will compare
the properties of several nearby Herbig stars 
which have been detected in X-rays. Some of these
detections are based on new or unpublished results, 
and we summarize these before proceeding.

\noindent{ {\bf Elias 1 (V892 Tau)}:}
This Herbig Ae star in the Taurus dark clouds was detected
in the {\em ROSAT} survey of ZP94. A companion located
4.1$''$ to the NE has been detected in high-resolution 
{\em VLA} radio observations (Skinner, Brown, \& 
Stewart 1993) and in the near-infrared (Leinert et al. 1997).
There is no definitive evidence so far of any companion
at closer separations. 

Giardino et al. (2004) have analyzed {\em Chandra} and
{\em XMM-Newton} observations of Elias 1. Variability was
detected in the  $\sim$18 ks {\em Chandra} light
curve, with impulsive changes in the count rate of   
a factor of two, suggestive of low-level flaring.
Analysis of the {\em Chandra} spectrum with a 
1T thermal plasma model gave a characteristic
temperature kT $\approx$ 2 keV and 
log L$_{X}$ (0.5 - 7.5 keV) = 30.2 ergs s$^{-1}$,
the latter value being comparable to earlier
{\em ROSAT} results (Table 5). The {\em XMM-Newton}
observation detected a large high-temperature flare  
that was attributed
to Elias 1, to within the {\em XMM-Newton} absolute
pointing accuracy of $\approx$4$''$. 

The detection of rapid X-ray variability and hot
plasma is a signature of magnetic activity. This is an
important result if the emission is intrinsic to
the Herbig star Elias 1 and not due to an as
yet undetetected lower mass companion. 
Because of the unusual X-ray properties of Elias 1,
further searches for a close companion are clearly
warranted.

\noindent{ {\bf HD 150193}:}
This Herbig Ae star was 
detected as a double X-ray
source in the 2.92 ks {\em Chandra} snapshot 
observation of FLG03,  shown in Figure 6. 
A  faint source (C) is located $\approx$1.36$''$
NE (PA = 50$^{\circ}$ $\pm$ 5$^{\circ}$) of the 
brighter X-ray peak (A). We use abbreviations
(A) and (C) to identify bright X-ray peak and the
faint source to its NE, consistent with the 
notation in FLG03.
Near IR observations also show a double source, with 
the bright IR source being 2.2 mag brighter at K band
and lying 1.1$''$ NE (PA = 56$^{\circ}$) of the 
faint IR source  (Pirzkal, Spillar,
\& Dyck 1997). The nearly identical separations 
and position angles suggest a one-to-one
correspondence between the two sources seen 
in the X-ray and IR images. However, because of
the close  separation there is some ambiguity
as to which source is the Herbig star.

Our measurements of the X-ray positions of the
two sources in the aspect-corrected {\em Chandra}
image (Fig. 6) indicate that the X-ray peak of the faint
northerly source lies closer to the optical 
position of HD 150193, with a position offset of
0.6$''$. Thus, the faint X-ray source (C) is 
most likely the Herbig star. To match the relative
geometry seen in the IR image, we then associate
the faint northern X-ray source (C) with the 
bright northern IR source. That is, the Herbig
star is faint in X-rays but bright in the near-IR.
The X-ray bright source to the south is then 
associated with the fainter IR companion to the south. 
Note that this interpretation is different than 
that given by FLG03, who 
associated the brighter southern X-ray source (A)
with the brighter northern IR source. In that
case, there is no {\em Chandra}  X-ray counterpart 
to the fainter southern IR source.
We estimate 17 counts for the faint northern X-ray
source (C) that we associate with HD 150193. The
PIMMS simulator\footnote{The Portable Interactive
Multi-Mission Simulator (PIMMS) is a software tool 
developed and maintained by HEASARC at NASA's GSFC.
For documentation, see 
http://heasarc.gsfc.nasa.gov/docs/software/tools.}
then gives 
log L$_{X}$ (0.5 - 7 keV) = 29.2 ergs s$^{-1}$
(Table 5).

\noindent{ {\bf HD 163296}:}
This Herbig star of spectral type A1Ve was observed 
for 10.7 ks  with the  {\em ROSAT} ~High Resolution
Imager (HRI) on 31 March 1995. 
The archived image (rh202032n00) shows a clear detection of
a source within 2$''$ of the optical position of
HD 163296. Using PIMMS, the HRI count rate of 0.012 c s$^{-1}$
gives an intrinsic luminosity 
log L$_{X}$ (0.5 - 7 keV)  = 29.8 ergs s$^{-1}$ (Table 5). 
This value  of L$_{X}$ is very similar to that of
other Herbig A0-1e stars such as AB Aur.

Close scrutiny of the apparent X-ray detection of 
HD 163296 is warranted because this star emits strongly
in the UV (log T$_{eff}$ (K) = 3.97, AWD98) and 
the HRI is known to be sensitive to UV radiation.
This UV sensitivity was apparent in HRI calibration
observations of the A0V star Vega (V = 0.03 mag), 
which gave a count rate of 0.10 c s$^{-1}$ and a
soft photon pulse height channel distribution. This
count rate was much larger than expected based on
HRI UV leak predictions (David et al. 1999; see also
Bergh\"{o}fer, Schmitt, \& H\"{u}nsch 1999 and
Barbera et al. 2000). A previous analysis
concluded that the HRI count rate of HD 163296
was compatible with that expected for  UV
leaks (Grady et al. 2000). However, we have reexamined
the HRI data and reach a different conclusion,
as discussed below.
 
If we make the worst-case assumption that
the Vega count rate was due entirely to UV
leaks, then a calculation based on the V
magnitude of HD 163296 (V = 6.85; Th\'{e} et al.
1985) gives a predicted
count rate due to UV contamination of 
$\approx$1.9 $\times$ 10$^{-4}$ c s$^{-1}$,
which is $\sim$60 times less than observed.
An alternative calculation using the HRI UV
leak calibration derived by Bergh\"{o}fer,
Schmitt, \& H\"{u}nsch (1999) and U = 6.99 mag
for HD 163296 gives a predicted rate due to
UV leaks of $\approx$1.3 $\times$ 10$^{-5}$ c s$^{-1}$,
or a factor of $\sim$900 less than observed.
In addition, we find that the pulse height channel
distribution of photons detected from HD 163296
by HRI is significantly harder than that of 
Vega. Based on these UV leak count rate predictions and
photon hardness considerations, we conclude that
the HRI observation yielded a valid X-ray detection
of HD 163296.  We thus include HD 163296 in our
analysis of Herbig star X-ray sources below.

\noindent{ {\bf R CrA and T CrA}: }
These two stars are of interest because of their
proximity in the nearby Corona Australis dark cloud.
We assume a distance of 150 pc (FM84), but values 
in the literature range from 130 pc (Hillenbrand et al. 1992) 
to 170 pc (Knude \& H$\o$g 1998). R CrA has a variable spectrum
(A1e - F7e) and T CrA is classified as F0-5e
(FM84; Hillenbrand et al. 1992). Because
of its later Fe spectral type, T CrA is not strictly
a member of the Herbig star class but is often included
in Herbig star catalogs because of its similar
properties. Neither of these two stars was detected
in the {\em ROSAT} survey of ZP94. We have analyzed
an archived 19.7 ks {\em Chandra } observation of the CrA
dark cloud obtained on 7 October 2000 (ObsId = 19)
which shows a clear detection of R CrA and a 
probable detection of T CrA (Fig. 7). We obtain
68 $\pm$ 8 counts for R CrA and 4 $\pm$ 2 counts
for T CrA in the  0.5 - 7 keV range.  The emission
from T CrA is faint but very likely real because of
the  low ACIS-I background ($<$1 count) and 
the good agreement between the stellar and
X-ray positions (offset = 0.97$''$ after applying the
recommended aspect correction).
Using the PIMMS simulator, we obtain intrinsic
luminosities log L$_{X}$ (0.5 - 7 keV) = 29.1  for
R CrA  and  log L$_{X}$ = 27.9 (ergs s$^{-1}$)
for T CrA (Table 5). Since T CrA is a marginal
detection, its L$_{X}$ could be interpreted
more conservatively as an upper limit.

\section{X-ray Versus Bolometric Luminosities in Nearby Herbig Stars} 
Sensitive X-ray observations of several different
star-forming regions have revealed a statistically
significant correlation between the X-ray and
bolometric luminosities of T Tauri stars. This
correlation has been confirmed in  {\em Chandra}
observations of the young cluster IC 348 
(Preibisch \& Zinnecker 2002), the Orion Nebula 
Cluster (Feigelson et al. 2003), and  the 
embedded infrared cluster in NGC 2024 (Skinner,
Gagn\'{e}, and Belzer 2003). These studies 
consistently find a ratio 
log (L$_{X}$/L$_{bol}$) 
$\approx$ $-$3.75 $\pm$ 1.0, where the 
large  scatter is probably due in part
to X-ray flaring.
A correlation of L$_{X}$ with
stellar mass was also found in IC 348 and
the Orion Nebula Cluster. The physical 
origin of these correlations is not yet
known but they are thought to be fundamentally
linked to the X-ray emission process.

It would thus be of considerable interest to
determine if a relation between L$_{X}$ and
L$_{bol}$ also exists in Herbig stars, since
they are thought to be more massive counterparts
of CTTS. However, the observational issues 
are more complex in the case of Herbig stars.
Unlike T Tauri stars, there are no large
codistant samples of Herbig stars in nearby
clusters on which to base a statistical
study. Instead, Herbig stars are sparsely
distributed over many different star-forming
regions spanning a range of uncertain distances 
and ages. 

To make an initial comparison between 
L$_{X}$ and L$_{bol}$, we have identified
ten nearby Herbig stars (d $\ltsimeq$ 200 pc)
whose distances are relatively well-known
from {\em Hipparcos} parallaxes or by
association with dark clouds.
Their L$_{X}$ and L$_{bol}$ values are
listed in Table 5 and plotted in Figure 8,
normalized to the quoted distances.
All stars in this subsample have been detected 
in X-rays and
their L$_{X}$ values were determined
from recent {\em Chandra} observations (FLG03),
the {\em ROSAT} survey of ZP94, and our own
analysis of archived {\em Chandra} and
{\em ROSAT} data. Bolometric luminosities
were taken from the literature 
(AWD98; Berrilli et al. 1992). We note that
there are conflicting values of the luminosity
of Elias 1 in the literature (Berrilli et al. 1992;
Hillenbrand et al. 1992).  This subsample
consists of stars of spectral type
B9e or later and does not include
early B-type stars, which typically lie
beyond $\sim$500 pc and have very uncertain
distances.

Since the data are based on an incomplete 
distance-limited sample, they do not
answer the question of whether 
L$_{X}$ and L$_{bol}$ are correlated in
the larger population of Herbig stars,
of which more than 100 are presently known.
However, several interesting trends 
can be seen in Figure 8.
First, the  values
$-$6.2 $\leq$ log (L$_{X}$/L$_{bol}$) $\leq$ $-$4.7
are clearly smaller than that observed for TTS
and larger than that usually found in OB stars.
This suggests that if the X-ray emission is
intrinsic to the Herbig stars themselves 
(as opposed to a companion origin),
then the efficiency of the X-ray emission process
as gauged by L$_{X}$/L$_{bol}$ ratios 
is lower than in TTS. 
Second, the range in log L$_{X}$ is consistent
with that found for TTS. 
Third, even for the
restricted range in  L$_{bol}$ of this 
subsample, the scatter in  L$_{X}$ is
large ($\pm$1.0 dex) but is still comparable 
to the scatter observed in TTS.

Figure 8 also includes two  emission-line stars of
later Fe spectral type  for
which {\em Chandra} data exist. These  are
the weakly-detected star T CrA (F0-5e) and
the spectroscopic binary AK Sco (F5$+$F5IVe; FLG03),
which consists of two nearly identical Fe
stars of masses M$_{*}$ $\approx$ 1.4 M$_{\odot}$
in a 13.6 day orbit (Alencar et al. 2003).
Even though these stars are not members of 
the Herbig Ae/Be class, they may be slightly
lower mass analogs. Despite the nearly identical
spectral types and  similar L$_{bol}$ values for these
two stars, there is a remarkable difference in
their L$_{X}$, with the binary AK Sco being more than
an order of magnitude brighter in X-rays. This
comparison shows that the presence of a companion
has by some means greatly increased the X-ray output
of AK Sco. It is tempting to speculate that the 
much fainter X-ray emission of T CrA may be more
representative of young single Fe-type stars.
However, there are hints that even T CrA may
be a binary. Even though no companion was
detected in  K-band speckle images 
(Ghez et al. 1997),  the position spectra of
Takami, Bailey, \& Chrysostomou (2003) suggest
that a faint companion could be present at a 
separation $>$0.14$''$.
Since this lower limit is larger than the
limit imposed by the Ghez et al K-band speckle images, 
Takami et al. concluded that if a companion is present it
must be fainter than K = 10.5 mag.

\section{Comparison With Classical Be Stars}
Classical Be stars are B-type stars on or near
the main sequence that exhibit line emission
above the photospheric spectrum, as  
recently  reviewed by Porter and 
Rivinius (2003). The evolutionary status of
classical Be stars is not well-understood.
However, they are rapid
rotators, as are Herbig Ae/Be stars. This has
led to the suggestion that Herbig stars might
be the evolutionary precursors of classical
Be stars (Finkenzeller 1985; Palla 1991).

Figure 8 shows the approximate region in the 
(L$_{bol}$, L$_{X}$) plane occupied by
classical main-sequence BVe stars, based on a 
search of the {\em ROSAT} all-sky survey (RASS)
catalog of optically bright OB stars compiled
by Bergh\"{o}fer, Schmitt, \& Cassinelli (1996).
Most BVe stars detected in the RASS had
29.5 $<$ log L$_{X}$ $\leq$ 31.0  ergs s$^{-1}$,
and a few undetected stars had upper limits
as low as log L$_{X}$ $\leq$ 28.5  ergs s$^{-1}$.

As can be seen in Figure 8, the range of 
L$_{X}$ values for classical BVe stars is
very similar to that of our Herbig star 
sample, but the L$_{X}$/L$_{bol}$ ratios are
clearly smaller for BVe stars. For BVe
stars detected in the RASS, typical values
are log (L$_{X}$/L$_{bol}$) = $-$6.4 $\pm$ 0.5,
whereas the Herbig star sample has
log (L$_{X}$/L$_{bol}$) = $-$5.4$^{+0.8}_{-1.2}$.
This comparison indicates that if the X-ray
emission is intrinsic to the Herbig stars and
if they do eventually evolve into classical Be
stars, then their L$_{X}$/L$_{bol}$ ratios
must decline by roughly an order of magnitude
after reaching the main sequence. Since no
significant increase in L$_{bol}$ is expected
as a Herbig star evolves onto the main sequence,
the above evolutionary scenario would imply
that L$_{X}$ must drop sharply. A dramatic
decrease in L$_{X}$ during the early main
sequence phase is predicted by the shear-induced 
dynamo model, as discussed in more detail below  
(Sec. 7.3).

\section{X-ray Emission Mechanisms}

\subsection{General Constraints on Emission Mechanisms}
An analysis of several different possible X-ray emission
mechanisms for HD 104237 was given by SY96 based on 
{\em ASCA} spectra. They examined the feasibility of
wind shock and accretion shock models as well as 
models involving magnetic confinement (coronae;
wind-fed magnetospheres). They argued that shock
models and magnetically-confined wind models could
not explain the hotter plasma seen in the {\em ASCA}
spectrum and concluded that the X-rays  most likely
originate in a magnetically confined region such 
as a corona. However, {\em ASCA}'s moderate
angular resolution could not distinguish between 
X-ray emission from HD 104237 and other sources
within $\approx$30$''$, as can now be done (Fig. 2).

The spectral properties derived
from the new {\em XMM-Newton} data are very similar to
those obtained with {\em ASCA} and strengthen the 
previous conclusions of SY96. In particular, the existence 
of hot plasma at or above kT $\approx$ 3 keV ($\sim$35 MK) is 
now substantiated by the detection of high-temperature
features such as the Fe K shell complex in the EPIC spectra.
These high
temperatures are incompatible with the predicted values
of kT $<$ 1 keV from wind shock and accretion shock models,
assuming a wind speed $v_{\infty}$ $\sim$ 500 km s$^{-1}$
and free-fall speed $v_{ff}$ $\sim$ 560 km s$^{-1}$ for
a Herbig Ae star of $\sim$2 M$_{\odot}$ (SY96). Shock 
models can account for the hotter plasma only if the
actual wind speeds or infall speeds exceed the above
estimates by a factor of two or more.

On the other hand, plasma temperatures of
kT $\approx$ 3 keV are quite typical of magnetically active
stars, including T Tauri stars. This value is well above
the hydrogen escape temperature T$_{esc}$ $\approx$ 13 MK (assuming
M$_{*}$ =  2.3 M$_{\odot}$, R$_{*}$ = 2.7 R$_{\odot}$; Table 1)
and it is therefore likely
that the emission arises in magnetically-confined plasma.
This could occur either in the corona of an as yet
undetected late-type companion or perhaps in
the Herbig star itself, as discussed further below.

\subsection{X-ray Emission from Late-Type Companions}
The companion hypothesis attributes the X-ray emission
to a fainter low-mass companion star, rather than the
Herbig star itself. This avoids the need to postulate
magnetic fields in an intermediate mass star that is
presumed to be non-convective, thus circumventing the 
contradiction with models based on the solar paradigm
which associate magnetic fields with convection.
Assuming that the companions are coeval with the
Herbig star primary, they would most likely be 
T Tauri stars or perhaps optically-obscured
class I infrared sources in the case of the
youngest Herbig stars.

As already noted, fainter sources have been found
in the immediate vicinity of HD 104237. These 
include the weak X-ray sources B and C detected
by {\em Chandra} (Fig. 2) and the report of a
faint star at an offset of 1.365$''$ and PA =
254.6$^{\circ}$ $\pm$ 0.35$^{\circ}$ detected in
{\em HST STIS} images (G04). 
However, it is obvious from the {\em Chandra} image
in Figure 2 that the faint sources B and C are
not the origin of the strong X-ray emission
detected at or near the HD 104237 optical 
position since they are clearly visible as
secondary peaks lying $\approx$4$''$ - 6$''$ to 
the west.

It is also unlikely that the faint star at a
separation of 1.365$''$ detected by {\em HST} is
the origin of the X-ray emission  associated
with HD 104237, for three reasons. First, an 
offset of 1.365$''$ between the X-ray and optical
positions would have been discerned at the positional
accuracy of the {\em Chandra} image, which is 
better than 0.6$''$ (Sec.  3.0.1). Second, the
position angle of this faint star relative to
HD 104237 is inconsistent with the PA of the
X-ray peak relative to the HD 104237 optical 
position. The centroid and peak-pixel  X-ray 
positions lie $\approx$0.4$''$ - 0.5$''$ to the 
north of the {\em Hipparcos} optical position
(Fig. 2), while the faint M-type companion lies 1.365$''$
to the W/SW at PA = 254.59$^{\circ}$ 
Third, the X-ray luminosity
log L$_{X}$ = 30.5 $\pm$ 0.1 (ergs s$^{-1}$) of 
the bright source detected here is at least an order of 
magnitude larger than typically observed for
M3-4 TTS (e.g. Preibisch \& Zinnecker 2002).
 
Since the M-type star at 1.365$''$ separation is not
the primary X-ray source, the X-ray emission must
arise from HD 104237 itself or from an as yet undetected
companion at a separation $r$ $<$ 1.365$''$. Our
comparison of the X-ray and optical positions (Fig. 2) 
indicates that the dominant X-ray source lies within 
0.6$''$ of the Herbig star optical position. 
{\em HST} results provide an additional constraint
by excluding a T Tauri companion at separations 
0.05$''$ $<$ $r$ $<$ 1.0$''$ (G04). Taken together,
the X-ray and {\em HST} observations imply that if
the X-ray emission is due to an as yet undetected
T Tauri companion, it must lie very close to 
HD 104237 at a separation  $r$ $\leq$  0.05$''$.

As Figure 7 shows, the X-ray luminosities determined
for nearby Herbig stars are within the range observed  
for T Tauri stars. Also, the scatter in log L$_{X}$ for 
this subsample of Herbig stars is $\pm$1.0 dex, which
is comparable to that of TTS. Thus, from the standpoint
of X-ray luminosities alone, one cannot exclude close T Tauri 
companions as the origin of the X-ray emission from
Herbig stars. Assuming that all of the X-ray emission
attributed to HD 104237 arises in an as yet undetected
T Tauri companion,
then the canonical relation log (L$_{X}$/L$_{bol}$)
$\approx$ $-$3.75 for TTS yields a bolometric 
luminosity log L$_{bol}$ $\approx$ 34.25, or
L$_{bol}$ $\approx$ 4 - 5 L$_{\odot}$ for the
putative companion. However, this value is 
highly uncertain because of the scatter in the
L$_{X}$ versus  L$_{bol}$ relation for TTS.

\subsection{Intrinsic X-ray Emission from Herbig Stars}
The alternative to the companion hypothesis is that
some, or all, of the X-ray emission is due to the 
Herbig star itself. Assuming that magnetic confinement
is necessary to explain the observed high-temperature 
plasma, then the emission could very well arise in
a corona. Another possibility
is a magnetically-confined wind shock, as has been proposed
to explain the X-ray emission of some  magnetic Ap-Bp stars
such as IQ Aur (Babel \& Montmerle 1997). However,
this model gives temperature predictions kT $<$ 1 keV that 
are similar to non-magnetic shock models, assuming wind
speeds v$_{\infty}$ $\approx$ 400 - 600 km s$^{-1}$ 
typical of Herbig Ae stars. Thus, this model cannot 
explain the high-temperature plasma at kT $\gtsimeq$ 3 keV.

In late-type stars, an outer convection zone is thought to
be necessary to support magnetic (e.g. coronal) activity via a
solar-like dynamo. The region in the HR diagram where surface
convection zones first appear is somewhat uncertain, as 
recently reviewed by  Favata \& Micela (2003). However, the
{\em ROSAT} detection of intrinsic X-ray emission from Altair
(A7IV-V) (Schmitt 1997) provides compelling evidence that 
surface convection is already present in some late A-type stars.
The recent classification of HD 104237 as A7.5Ve - A8Ve based
on {\em HST} observations (G04) thus raises the interesting
possibility that some  of its X-ray emission might be 
associated with surface convection. However, the evolutionary 
tracks of Siess et al. (2000) suggest that 
a surface convection zone 
in HD 104237 would be quite thin, amounting to $\approx$0.9\% of
the stellar radius (assuming T$_{eff}$ = 7300 K, log [L$_{*}$/L$_{\odot}$] 
= 1.42). It is not obvious why such a thin convection zone would
give rise to an  X-ray luminosity that is $\sim$1000 times
greater than that of Altair, which has a similar A7V spectral type.
It may thus be that other factors, perhaps associated with youth,
lead to elevated X-ray emission levels of Herbig stars - as
discussed below.

It has been suggested that coronal X-ray emission could  
be sustained in a young rapidly-rotating
Herbig star via a magnetic field set up by non-solar dynamo action
that is powered by rotational shear energy. 
This possibility was explored by Vigneron et al. (1990)
in the theoretical framework of sheared stratified fluids.
A later development of this model by Tout \& Pringle 
(1995, hereafter TP95) provides a  quantitative prediction 
of the X-ray luminosity versus age, L$_{X}$(t), but makes
no temperature predictions. In the TP95 formulation, 
\begin{equation}
{\rm L_{X}(t)~ = ~ L_{Xo} \left[1 + \frac{t}{t_{0}}\right]^{-3} }
\end{equation}
In the above, the initial X-ray luminosity L$_{\rm Xo}$
depends on the stellar mass M$_{*}$, radius R$_{*}$,
change in the star's angular velocity between its center
and surface ($\Delta\Omega_{0}$), breakup velocity ($\Omega_{k}$),
and two dimensionless parameters of the theory $\epsilon$
and $\gamma$ according to (eq. 4.4 of TP95)

\begin{equation}
{\rm \frac{L_{Xo}}{L_{\odot}} ~\sim~ 2.63 \times 10^{-3}
\left[\frac{\epsilon}{10^{-3}}\right] \left[\frac{\gamma}{3 \times 10^{-5}}\right]^{3}
\left[\frac{\Delta\Omega_{0}}{\Omega_{k}}\right]^{3} \left[\frac{M_{*}}{M_{\odot}}\right]^{2.5}
\left[\frac{R_{*}}{R_{\odot}}\right]^{-2.5}.    }
\end{equation}

For computational purposes, it is assumed that  $\Delta\Omega_{0}$/$\Omega_{k}$
$\sim$ 1 (TP95). The parameters $\epsilon$ and $\gamma$ are not well-determined
empirically, with $\epsilon$ being the fraction of
magnetic flux that is
dissipated at the stellar surface to heat coronal
plasma and $\gamma$  the efficiency of magnetic
field generation. The decay timescale {\rm t$_{0}$ }
also depends on  $\gamma$,  M$_{*}$, and R$_{*}$.
For fiducial values $\epsilon$ $\sim$ 10$^{-3}$
and $\gamma$ $\sim$ 3 $\times$ 10$^{-5}$ (TP95),
the decay timescale for Herbig Ae stars of mass
M$_{*}$ $\approx$ 2 - 3 M$_{\odot}$ of interest here is
{\rm t$_{0}$ } $\sim$1 - 2 My. Thus, this mechanism
could sustain magnetic activity in a Herbig Ae star
for a few million years, but drops off rapidly as
t$^{-3}$ thereafter.

A  comparison of the predicted L$_{X}$ from the above
model with that measured for HD 104237 by {\em ASCA} was made by
SY96. They assumed an earlier A4IVe spectral type, 
mass M$_{*}$ = 2.1 M$_{\odot}$, radius
R$_{*}$ = 2.6 R$_{\odot}$, and age t $\sim$ 2 My.
For these parameters, the model underestimated the
observationally-determined L$_{X}$ by a factor of
$\sim$4. But the more recent {\em HST} data suggest a later
A8Ve spectral type, T$_{eff}$ = 7300 K,  stellar
luminosity L$_{*}$ = 26.3 L$_{\odot}$, and an age
t $\sim$ 5 My (G04). Using these parameters, the 
evolutionary tracks of Siess et al.
(2000) give M$_{*}$ = 2.14 M$_{\odot}$ and
R$_{*}$ = 3.0 R$_{\odot}$. In that case, the 
decay timescale is t$_{0}$ $\sim$ 2.2 My, and
the model underestimates the X-ray luminosity (Table 1)  by
a factor of $\sim$25. The above calculations show that
uncertainties in ages, masses, and  theoretical
parameters such as $\epsilon$ and $\gamma$ currently limit
observational tests of the TP95 model to 
order-of-magnitude comparisons. 

We have made additional comparisons of the predicted
L$_{X}$ from the TP95 model with the observed values 
listed for the other Herbig stars in Table 5. In these
comparisons we have used stellar parameters
given in the literature (i.e. AWD98) as well as our
own estimates from the Siess et al. tracks, along
with the fiducial values of $\epsilon$ and
$\gamma$ given above.  Elias 1 was excluded from this
comparison since it was not included in the AWD98 
{\em Hipparcos}-based study
and there are significant differences in
reported values for its stellar luminosity and 
radius in the literature (e.g. Berrilli et al. 1992; 
Hillenbrand et al. 1992).  

The model gives very good agreement for AB Aur
and HD 97048, and it may be relevant that speckle
techniques have so far failed to find any evidence
for close companions around these two stars 
(Leinert, Richichi, \& Haas 1997; Ghez et al. 1997).
However, the model tends to underestimate L$_{X}$
by as much as an order-of-magnitude in other cases,
as already noted above  for HD 104237.  
The discrepancy could be even larger for the
B9Ve star HD 141569 if its age is t $>$ 10 My 
(AWD98). However, a younger age
t $\sim$ 5 $\pm$ 3 My has been inferred  
based on its association with two
young M-type common proper motion
companions (Weinberger et al. 2000). If
this younger age is correct, then the 
model prediction is within an order of 
magnitude of the observed value for HD 141569.

\section{Summary}
The main results of this study are the following:
\begin{enumerate}

\item New X-ray spectra of the Herbig Ae star HD 104237
obtained with {\em XMM-Newton} provide the first clear
detection of emission lines in this star, including
Si XIII, Ca XIX, and the Fe K$\alpha$ complex. The X-ray emission 
arises in an absorbed  multi-temperature thermal plasma 
with cool (kT $<$ 1 keV) and hot (kT $\gtsimeq$ 3 keV)
components. The inferred X-ray absorption is equivalent 
to an extinction A$_{V}$ $\approx$ 0.3 - 1.0 mag, with 
best-fit models converging toward the high end of this range.
This range is consistent with that determined from optical
studies.

\item The presence of hot plasma at kT $\gtsimeq$ 3 keV
is not consistent with shock model predictions, but is
typical of values observed in magnetically-active 
PMS stars such as T Tauri stars. We conclude that the
emission arises in magnetically-confined plasma,
substantiating earlier {\em ASCA} results. 
The {\em XMM-Newton} light curve shows signs
of slow variability at the $\approx$15\% level on 
a timescale of $\sim$hours, and a comparison of
two  {\em Chandra} observations taken eight months
apart also suggests that low-level variability may
be present. More persistent time monitoring
over longer timespans is needed to confirm the 
suspected variability and constrain the timescale(s).
If short-term ($\sim$hours) variability is confirmed,
it would lend further support to a magnetic
interpretation for the X-ray emission.
The precise nature and location of 
the magnetically-confined region is not yet known,
but a corona around the Herbig star (or an as yet
unseen close companion) is a likely possibility.

\item Archived {\em Chandra} images show that the X-ray
emission peak is offset by $<$0.6$''$ from the stellar
optical position, which equates to a projected separation
$<$70 AU at d = 116 pc. This small offset places stringent
limits on the angular separation of a putative late-type
X-ray companion, leaving open the possibility that some
(or all) of the X-ray emission arises in the Herbig star
itself.

\item The X-ray luminosities of a sample of ten Herbig
Ae/Be stars with reliable distances are in the range
log L$_{X}$ = 29.9 $\pm$ 1.0 (ergs s$^{-1}$), which 
overlaps the range observed for T Tauri stars. Thus,
X-ray emission from unseen T Tauri companions is not
ruled out on the basis of L$_{X}$ considerations. For this
distance-limited Herbig star sample,  we obtain a mean ratio 
log (L$_{X}$/L$_{bol}$) = $-$5.4$^{+0.8}_{-1.2}$, which is
less than for T Tauri stars and greater than for classical
BVe stars. If the hypothesis that Herbig stars evolve 
into classical BVe stars is true, then the L$_{X}$ 
values of Herbig stars must decrease dramatically
after reaching the main sequence, and such a decrease
is predicted by shear dynamo models.

\item The shear dynamo model of Tout \& Pringle (1995) 
provides an alternative
to the companion hypothesis for explaining X-ray emission in
Herbig stars. A comparison of the predictions of this model
against known  L$_{X}$ values for a sample of nine Herbig
stars shows very good agreement in two cases (AB Aur and
HD 97048). However, the model tends to underestimate
L$_{X}$ in all other cases by as much as an order of
magnitude.  Several factors could account for
this disagreement including uncertain ages and masses,
poorly-known model parameters that characterize 
magnetic processes, and excess X-ray emission from
unseen lower mass companions.

\end{enumerate}

This study strongly suggests that the X-ray emission 
detected within 0.6$''$ of the HD 104237 optical position 
is of magnetic origin. The question that remains is 
whether the emission is coming from the Herbig star itself,
or from an as yet undetected companion. The existing
{\em Chandra} images show that if the emission is due to
a companion, then its separation is $r$ $<$ 1$''$ from HD 104237.
Recent {\em HST} images (G04) place a tighter constraint of
$r$ $<$0.05$''$ on the separation of a T Tauri companion.
Such a close companion, if present, cannot be spatially resolved with
existing  X-ray telescopes.  Even so, further progress on the
companion issue may nevertheless be possible at X-ray 
wavelengths. For example, high-quality images and spectra
comparable to those of HD 104237 could be obtained for a
much larger sample of Herbig stars, selected from the more than
one hundred known and candidate members of the class
(Th\'{e}, de Winter, \& P\'{e}rez 1994). 
A more extensive X-ray data base
would provide more complete information on plasma properties
(kT, EM distribution), the X-ray luminosity function, and 
positional offsets. This information could then be used in
statistical comparisons  to determine if the X-ray
properties of T Tauri stars are significantly different from
Herbig Ae/Be stars.
\acknowledgments

This work was supported by NASA grant NAG5-10326 
and was based on observations obtained with 
{\em XMM-Newton}, an ESA science mission with instruments and
contributions directly funded by ESA member states
and the USA (NASA). Paul Scherrer Institut is financially supported for
X-ray astronomy by the Swiss National Science Foundation.
We thank members of the {\em XMM-Newton}
HEASARC (NASA/GSFC) and VILSPA support teams for their
assistance. This research has made use of the SIMBAD
astronomical database operated by CDS at Strasbourg,
France. 

%
%
%
\clearpage
%
%
\begin{deluxetable}{lll}
\tabletypesize{\scriptsize}
\tablewidth{0pc}
\tablecaption{Stellar Properties of HD 104237 \label{tbl-1}}
\tablehead{
\colhead{Parameter}      &
\colhead{Value  }  &
\colhead{Notes\tablenotemark{a}  }     
}
\startdata
RA (J2000)                           & 12$^h$ 00$^m$ 05.0846$^s$        & (1)  \nl
Decl. (J2000)                        & $-$78$^{\circ}$ 11$'$ 34.564$''$ & (1)  \nl
Sp. Type                             & A4IVe; A7IVe; A7.5Ve - A8Ve      & (2),(3),(4),(5) \nl
d (pc)                               & 116$^{+8}_{-7}$                  & (1) \nl 
V (mag.)                             & 6.6                              & (6)  \nl
A$_{V}$ (mag)                        & 0.3 - 0.9                        & (7),(8) \nl
J, H, K (mag)                        & 5.8, 5.2, 4.6                    & (9) \nl
Mass (M$_{\odot}$)                   & 2.3                              & (7),(10) \nl
log L$_{*}$ (L$_{\odot}$)            & 1.55$^{+0.06}_{-0.05}$           & (7),(10) \nl
log L$_{X}$ (ergs s$^{-1}$)          & 30.5$\pm$ 0.1                     & (11)
\enddata
\tablenotetext{a} {Refs. and Notes:
(1) {\em Hipparcos} position from Perryman et al. (1997), corrected for proper
    motion to epoch J2000.0~
(2) Hu et al. (1991)~
(3) Eggen (1995)~
(4) Brown et al. (1997)~
(5) Grady et al. (2004)~
(6) SIMBAD~
(7) van den Ancker et al. (1998)~
(8) Malfait et al. (1998)~
(9) 2MASS Point Source Catalog~
(10) The quoted mass and L$_{*}$ are from van den Ancker et al. (1998). The recent
     {\em HST} study of Grady et al. (2004) obtained log L$_{*}$ (L$_{\odot}$) =
     1.42$^{+0.04}_{-0.07}$ and T$_{eff}$ $\approx$ 7300 K, for which the 
     evolutionary tracks of Siess et al. (2000) give M$_{*}$ = 2.1 M$_{\odot}$.~
(11) this work; quoted value is the unabsorbed X-ray luminosity (0.5 - 7 keV) obtained by
     averaging the values obtained from 2T and 3T VAPEC models (Table 4)
}
\end{deluxetable}
\clearpage
%
\begin{deluxetable}{ll}
\tablewidth{0pc}
\tablecaption{XMM-Newton Observations of HD 104237 }
\tablehead{
\colhead{Parameter}      &
\colhead{Value  }     
}
\startdata
Start (UT)                        & 17 Feb 2002 15:50                                    \nl
Stop  (UT)                        & 17 Feb 2002 19:52                                    \nl
Usable Exposure (ks)              & 12.2 (PN), 14.4 (per MOS)                             \nl
ObsId / Revolution                & 0059760101 / 402 \nl
EPIC Mode                         & Imaging - Full Window                                \nl 
Optical Filter                    & Thick           
\enddata
\end{deluxetable}

\clearpage
\begin{deluxetable}{llllll}
\tabletypesize{\scriptsize}
\tablewidth{0pt}
\tablecaption{XMM-Newton Sources\tablenotemark{a} }
\tablehead{
\colhead{No.}      &
\colhead{Name}     &
\colhead{R.A.}       &
\colhead{Decl.}    &
\colhead{Net Counts}   &
\colhead{Identification}      \\
\colhead{     }    &
\colhead{     }    &
\colhead{(J2000)     }    &
\colhead{(J2000)     }    &
\colhead{            }    &
\colhead{     }    
}
\startdata
1  & J115646.4$-$781325  &11 56 46.47 &  -78 13 25.0 & 42 $\pm$ 12 &  ...          \nl
2  & J115908.7$-$781234  &11 59 08.76 &  -78 12 34.6 & 72 $\pm$ 14 &  2M 11590798-7812322 \nl
3  & J115913.6$-$782409  &11 59 13.62 &  -78 24 09.7 & 54 $\pm$ 13 &  ...         \nl
4  & J115932.3$-$782223  &11 59 32.37 &  -78 22 23.1 & 90 $\pm$ 14 &  ...         \nl
5  & J115943.4$-$775838  &11 59 43.41 &  -77 58 38.4 & 66 $\pm$ 14 &  ...         \nl
6* & J115948.0$-$781147  &11 59 48.06 &  -78 11 47.4 & 46 $\pm$ 12 &  2M 11594807-7811450; CPD$-$77$^{\circ}$ 773 \nl 
7* & J120005.9$-$781136  &12 00 05.93 &  -78 11 36.6 &7167 $\pm$ 81&  2M 12000511-7811346; HD 104237 \nl 
8* & J120048.3$-$781107  &12 00 48.33 &  -78 11 07.6 & 47 $\pm$ 12 &  ... \nl 
9* & J120049.6$-$781001  &12 00 49.63 &  -78 10 01.9 & 57 $\pm$ 18 &  2M 12004942-7809574; CPD$-$77$^{\circ}$ 775 \nl
10 & J120054.9$-$782030  &12 00 54.92 &  -78 20 30.7 & 38 $\pm$ 11 &  2M 12005517-7820296 \nl
11 & J120059.8$-$781658  &12 00 59.85 &  -78 16 58.8 &122 $\pm$ 32 &  ... \nl
12*& J120118.6$-$780254  &12 01 18.60 &  -78 02 54.3 & 56 $\pm$ 17 &  2M 12011809-7802522 \nl
13*& J120119.9$-$775947  &12 01 19.97 &  -77 59 47.5 &217 $\pm$ 33 &  2M 12012042-7759478 \nl
14 & J120135.7$-$780428  &12 01 35.75 &  -78 04 28.9 & 89 $\pm$ 21 &  ... \nl
15 & J120137.9$-$781832  &12 01 37.95 &  -78 18 32.2 & 38 $\pm$ 11 &  2M 12013691-7818345 \nl
16 & J120139.0$-$782120  &12 01 39.04 &  -78 21 20.2 & 67 $\pm$ 20 &  ... \nl
17 & J120153.2$-$781842  &12 01 53.25 &  -78 18 42.2 & 94 $\pm$ 14 &  2M 12015251-7818413 \nl
18 & J120156.5$-$780605  &12 01 56.57 &  -78 06 05.3 & 38 $\pm$ 11 &  ... \nl
19*& J120235.0$-$781059  &12 02 34.96 &  -78 10 59.3 & 92 $\pm$ 21 &  ... \nl
\enddata
\tablenotetext{a}{
Sources are included in this table if they were detected in PN and in
one or both MOS. An asterisk in column (1) indicates that additional
source notes follow below.
Net counts are from EPIC-PN pipeline processing source list, and are
background-subtracted and PSF-corrected
in the 0.5 - 7.5 keV range.
Candidate infrared and optical identifications are listed if
they lie within 5$''$ of the X-ray position. Infrared identifications
are from the 2MASS (2M) all-sky Point Source Catalog, and optical
identifications are from SIMBAD. \\
NOTES: \\
6. The K0 star CPD$-$77$^{\circ}$ 773 is offset by 2.4$''$ from the
X-ray position. Source lies near CCD gap; net counts may be
underestimated. \\
7.  Counts include nearby sources B,C,D, E (Fig. 2). \\
8.  Not detected by pipeline processing, but 
    visible in PN and MOS. Position is from MOS. \\
9.  The multiple star CPD$-$77$^{\circ}$ 775 is offset by 4.7$''$
    from the X-ray position. \\
12. Visible in PN but only weakly detected in MOS2. \\
13. Possible double or extended X-ray source. Peak position is from MOS. \\ 
19. Source lies near CCD gap; net counts uncertain. 
} \\
\end{deluxetable}
\clearpage
\begin{deluxetable}{lll}
\tabletypesize{\scriptsize}
\tablewidth{0pc}
\tablecaption{{\em XMM-Newton} Spectral Fit Results for HD 104237 
   \label{tbl-1}}
\tablehead{
\colhead{Parameter}      &
\colhead{2T VAPEC  }        &
\colhead{3T VAPEC  }
}
\startdata
N$_{\rm H}$ (10$^{21}$ cm$^{-2}$) & 0.78 [0.53 - 1.06]    & 2.25 [1.20 - 3.10] \nl
kT$_{1}$ (keV)                    & 0.58 [0.38 - 0.62] & 0.58 [0.46 - 0.63] \nl
EM$_{1}$ (10$^{52}$ cm$^{-3}$)& 4.46 [3.14 - 6.84]    & 6.71 [2.91 - 9.06]  \nl 
kT$_{2}$ (keV)                & 3.18 [2.78 - 3.47] & 2.92 [2.54 - 3.34] \nl
EM$_{2}$ (10$^{52}$ cm$^{-3}$)& 8.58 [7.34 - 10.1]  & 9.26  [7.02 - 10.5] \nl
kT$_{3}$ (keV)                & \nodata                & 0.15 [0.11 - 0.20] \nl
EM$_{3}$ (10$^{52}$ cm$^{-3}$)& \nodata            & 8.95 [1.79 - 12.0] \nl
Abundances                    & varied\tablenotemark{a} & varied\tablenotemark{b} \nl
$\chi^2$/dof                  & 221.0/217          & 208.8/215 \nl
$\chi^2_{red}$                & 1.02               & 0.96 \nl
F$_{\rm X}$ (10$^{-12}$ ergs cm$^{-2}$ s$^{-1}$) & 1.30 (1.53) & 1.32 (2.41) \nl
L$_{\rm X}$ (10$^{30}$  ergs s$^{-1}$)           & 2.46 & 3.88 \nl
\enddata
\tablecomments{
Based on  fits of the EPIC PN spectrum binned to a minimum of 
15 photons per bin (Fig. 5) with the VAPEC optically thin plasma 
model. The tabulated parameters
are absorption column density (N$_{\rm H}$), plasma temperature (kT),
and emission measure (EM).  Brackets enclose 90\% confidence intervals.
X-ray flux (F$_{\rm X}$) is the  observed (absorbed) value followed
in parentheses by the unabsorbed value in the 0.5 - 7.0 keV range.
X-ray luminosity (L$_{\rm X}$) is the unabsorbed value in the
0.5 - 7.0 keV range. 
The flux excludes counts in the proximity of faint nearby
sources detected in higher spatial resolution {\em Chandra} 
observations (Figs. 1 and 2).
A distance of 116 pc is assumed.} 
\tablenotetext{a} {Best-fit abundances relative to the solar values of
Anders \& Grevesse (1989) and 90\% confidence intervals were:
Ne = 2.3 [1.3 - 4.3], Mg = 2.0 [1.4 - 2.7], Si = 2.0 [1.2 - 3.2],
S = 0.7 [0.0 - 2.3], Ca = 9.9 [4.5 - 14.7], Fe = 0.8 [0.5 - 1.1].} 
\tablenotetext{b} {Best-fit abundances relative to the solar values
of Anders \& Grevesse (1989) and 90\% confidence intervals were:
Ne = 2.2 [1.3 - 4.9], Mg = 1.7 [1.0 - 4.0], Si = 1.6 [1.0 - 3.7],
S = 0.6 [0.0 - 2.1], Ca = 6.6 [1.5 - 13.1], Fe = 1.0 [0.7 - 1.8].}
\end{deluxetable}
\clearpage
\begin{deluxetable}{lllrccl}
\tabletypesize{\footnotesize}
\tablewidth{0pc}
\tablecaption{X-ray Detections of Nearby Herbig Ae/Be Stars 
              and Fe Stars \tablenotemark{a}     }
\tablehead{
\colhead{No.}      &
\colhead{Name  }  &
\colhead{Sp. Type }  &
\colhead{d  }  &
\colhead{log L$_{bol}$ }  &
\colhead{log L$_{X}$  }  &
\colhead{X-ray  }  \\
\colhead{ }      &
\colhead{  }  &
\colhead{ }  &
\colhead{(pc)  }  &
\colhead{ (ergs s$^{-1}$) }  &
\colhead{ (ergs s$^{-1}$) }  &
\colhead{refs.  }  
 } 
\startdata
1*  & HR 5999   & A5-7III/IVe  & 210$^{+50}_{-30}$ & 35.51 & 30.78 & (1)  \\
2   & HD 104237 & A4IVe        & 116$^{+8}_{-7}$   & 35.13 & 30.50 & (2,3,4)     \\
3*  & Elias 1   & A0-6e        & 140$^{+20}_{-20}$ & 35.16 & 30.28 & (1,5)         \\
4   & HD 141569 & B9.5Ve       &  99$^{+9}_{-8}$   & 34.93 & 30.10 & (2)     \\
5*  & HD 163296 & A1Ve         & 122$^{+17}_{-13}$ & 35.06 & 29.80 & (6)     \\
6*  & AB Aur    & A0Ve         & 144$^{+23}_{-17}$ & 35.26 & 29.50 & (1)      \\
7   & HD 100546 & B9Vne        & 103$^{+7}_{-6}$   & 35.09 & 29.40 & (2)     \\
8*  & HD 150193 & A1Ve         & 150$^{+50}_{-30}$ & 35.05 & 29.25 & (7)      \\
9*  & R CrA     & A1e-F7e (var)& 150$^{+30}_{-30}$ & 35.74 & 29.09 & (7)      \\
10* & HD 97048  & B9-A0e       & 180$^{+30}_{-20}$ & 35.19 & 28.96 & (1)     \\
11* & AK Sco    & F5$+$F5IVe   & 150$^{+40}_{-30}$ & 34.46 & 29.10 & (2)     \\
12* & T CrA     & F0-5e        & 150$^{+30}_{-30}$ & 34.58 & 27.88 & (7)    
\tablenotetext{a} {Hipparcos distances, spectral types, and L$_{bol}$  
  are from AWD98, except as follows. Distance of Elias 1 (= V892 Tau)
  is from Elias (1978), and distances of R and T CrA are from
  FM84. The L$_{bol}$ values of Elias 1, R and T CrA are from
  Berrilli et al. (1992). All  luminosities have been adjusted to the
  quoted distances and L$_{X}$ values are unabsorbed. 
  X-ray refs.: (1) ZP94 ({\em ROSAT}),
  (2) FLG03 ({\em Chandra}), (3) this work ({\em XMM}), 
  (4) SY96 ({\em ASCA}),
  (5) Giardino et al. 2004 ({\em Chandra}),
  (6) this work ({\em ROSAT} HRI archive), 
  (7) this work ({\em Chandra} archive). } 
\enddata
\tablecomments{ \\
1. L$_{X}$ is from ZP94,  adjusted to 210 pc. \\
3. Spectral type has been listed as A0 (FM84) or A6e (HB88). \\
5. L$_{X}$ is from PIMMS, using {\em ROSAT} HRI rate 0.012 c s$^{-1}$,
   N$_{H}$ = 5.5 $\times$ 10$^{20}$ cm$^{-2}$ (A$_{V}$ = 0.25 mag),
   and 1 keV Raymond-Smith (RS) plasma model. See text. \\
6. L$_{X}$ is from PIMMS, using {\em ROSAT} PSPC rate 8.85 $\times$
   10$^{-3}$ c s$^{-1}$ (ZP94), 
   N$_{H}$ = 1.1 $\times$ 10$^{21}$ cm$^{-2}$ (A$_{V}$ = 0.5 mag),
   and 1 keV RS plasma model. \\
8. L$_{X}$ is from PIMMS, using {\em Chandra} ACIS-I rate 5.82 $\times$
   10$^{-3}$ c s$^{-1}$, 
   N$_{H}$ = 3.5 $\times$ 10$^{21}$ cm$^{-2}$ (A$_{V}$ = 1.6 mag),
   and 1 keV RS plasma model. See text. \\
9. L$_{X}$ is from PIMMS, using {\em Chandra} ACIS-I rate 3.45 $\times$
   10$^{-3}$ c s$^{-1}$, 
   N$_{H}$ = 4.4 $\times$ 10$^{21}$ cm$^{-2}$ (A$_{V}$ = 2 mag),
   and 1 keV RS plasma model. See text. \\
10. L$_{X}$ is from PIMMS, using {\em ROSAT} PSPC rate 1.14 $\times$
   10$^{-3}$ c s$^{-1}$ (ZP94), 
   N$_{H}$ = 2.7 $\times$ 10$^{21}$ cm$^{-2}$ (A$_{V}$ = 1.2 mag),
   and 1 keV RS plasma model. \\
11. SB2 with 13.6 d orbital period (Alencar et al. 2003). \\
12. L$_{X}$ is from PIMMS, using {\em Chandra} ACIS-I rate 2.02 $\times$
   10$^{-4}$ c s$^{-1}$, 
   N$_{H}$ = (3.8 - 4.7) $\times$ 10$^{21}$ cm$^{-2}$ 
   (A$_{V}$ = 1.7 - 2.1 mag),
   and 1 keV RS plasma model. See text. 
}
\end{deluxetable}
\clearpage

\clearpage

\figcaption{Unsmoothed linearly-scaled {\em XMM-Newton} 
image of the region near HD 104237
obtained by combining photons from the MOS1 and MOS2 detectors
in the 0.3 - 7 keV range. The pixel size is 1.1$''$ and the coordinate
overlay is J2000.
Spectra and light curves of HD 104237 were extracted
using photons inside the solid circle of radius 18$''$ centered on
the star, excluding those photons that fell inside the dashed circles
surrounding the faint sources B,C,D, and E detected in higher-resolution
{\em Chandra} images (Fig. 2). {\em Chandra} positions of the four
faint sources are marked with crosses.
\label{fig1} }

\figcaption{Unsmoothed aspect-corrected {\em Chandra} ACIS-I image
of the region near HD 104237 obtained in a 2.86 ks observation
on 4 Feb. 2002 with HD 104237 located 1.7$'$ off-axis. The pixel
size is 0.49$''$ and the energy range is 0.3 - 7 keV, with a 
J2000 coordinate overlay and linear scaling. The asterisk at center marks the 
{\em Hipparcos} position of HD 104237 (Table 1). The {\em Hipparcos} position
is offset by 0.52$''$ from the X-ray {\em peak} position, which is (J2000.)  
RA = 12$^h$ 00$^m$ 05.062$^s$,
decl. = $-$78$^{\circ}$ 11$'$ 34.06$''$. The {\em Hipparcos} position
is offset by 0.4$''$ from the X-ray {\em centroid} position, which is
RA = 12$^h$ 00$^m$ 05.089$^s$,
decl. = $-$78$^{\circ}$ 11$'$ 34.16$''$.
Crosses mark the positions
of four fainter X-ray sources B,C,D, and E whose properties
are summarized in Tables 3 and 5 of FLG03.  
\label{fig2}}

\figcaption{Wavelet-smoothed {\em XMM-Newton} EPIC-PN image of the 
region surrounding HD 104237. The intensity scale is 
logarithmic and the energy range is 0.3 - 7 keV, with a 
J2000 coordinate overlay. The circled sources correspond
to objects detected in PN and MOS, as listed in Table 3.
\label{fig3}}

\figcaption{Background-subtracted  {\em XMM-Newton} X-ray light curve of 
HD 104237 using photons from the EPIC PN detector in the 0.5 - 5 keV
energy range and a bin size of 200 s (filled squares). The  light curve was 
extracted using the region shown in Figure 1, which excludes some photons
in the vicinity of faint sources B,C,D, and E. The background light
curve at bottom was extracted from a source-free region on the same
CCD as HD 104237 and is binned at 600 s intervals. The mean count
rate and standard deviation for the source light curve (top) is
0.25 $\pm$ 0.04 c s$^{-1}$ and the corresponding value for the 
background light curve (bottom) is 0.007 $\pm$ 0.004 s$^{-1}$.
Error bars are 1$\sigma$.
\label{fig4} }

\figcaption{Background-subtracted  {\em XMM-Newton} X-ray spectrum of
HD 104237 using photons from the EPIC PN detector and the extraction
region shown in Figure 1. The spectrum has been rebinned to a minimum
of 15 photons per energy bin. Solid line shows the best-fit 3T VAPEC
optically thin plasma model (Table 4). The best-fit 2T VAPEC model
(Table 4) is nearly identical to the 3T model shown, except
that the 2T model slightly underestimates the flux in the 0.5 - 0.6
keV range by $\approx$20\%. 
\label{fig5} }

\figcaption{Unsmoothed {\em Chandra} ACIS-I image of the 
Herbig Ae star HD 150193 in the 0.3 - 7 keV range with 
linear scaling and J2000 coordinates. The pixel size is
0.49$''$.  The cross marks the optical position
of HD 150193, and circled sources A and C correspond
to the bright X-ray peak and a fainter X-ray peak
to its NE. The X-ray position of the faint peak C is
RA = 16$h$ 40$m$ 17.94$s$, Decl. = $-$23$^{\circ}$ 53$'$ 44.64$''$.
which is offset by 0.6$''$ from the optical position, as 
compared to a 0.8$''$ offset for the bright peak A. The
faint peak C is identified with the Herbig star HD 150193.
\label{fig6} }

\figcaption{Unsmoothed archived {\em Chandra} ACIS-I image of
the Corona Australis cloud (ObsId = 19) in the  0.3 - 7 keV range. The
usable exposure is 19.7 ks and coordinate overlay is J2000.
The circled detection of R CrA has 68 $\pm$ 8 counts
and weak emission (4 $\pm$ 2 counts) is seen at the position
of T CrA.
\label{fig7} }

\figcaption{X-ray and bolometric luminosities of Herbig Ae/Be
stars (filled squares), whose numbers correspond to Table 5. 
Also shown are the Fe stars AK Sco and T Cra (open squares).
The stellar luminosity of Elias 1 is uncertain (Berrilli et
al. 1992; Hillenbrand et al. 1992).
The large solid polygon to the left encloses the region
occupied by T Tauri stars and  the sloping dashed
line shows the typical relation log (L$_{X}$/L$_{bol}$)
$=$ $-$3.75 for TTS. The solid polygon on the right
shows the analogous region for massive OB stars and
their canonical relation log (L$_{X}$/L$_{bol}$) $=$
$-$7.0.  The dotted polygon at right center encloses
the region occupied by classical main-sequence BVe stars.
The Herbig stars shown are (1) HR 5999, (2) HD 104237,
(3) Elias 1 (= V 892 Tau), (4) HD 141569, (5) HD 163296,
(6) AB Aur, (7) HD 100546, (8) HD 150193, (9) R CrA,
(10) HD 97048.
\label{fig8} }


\clearpage

\begin{figure}
\figurenum{1}
\epsscale{0.6}
\plotone{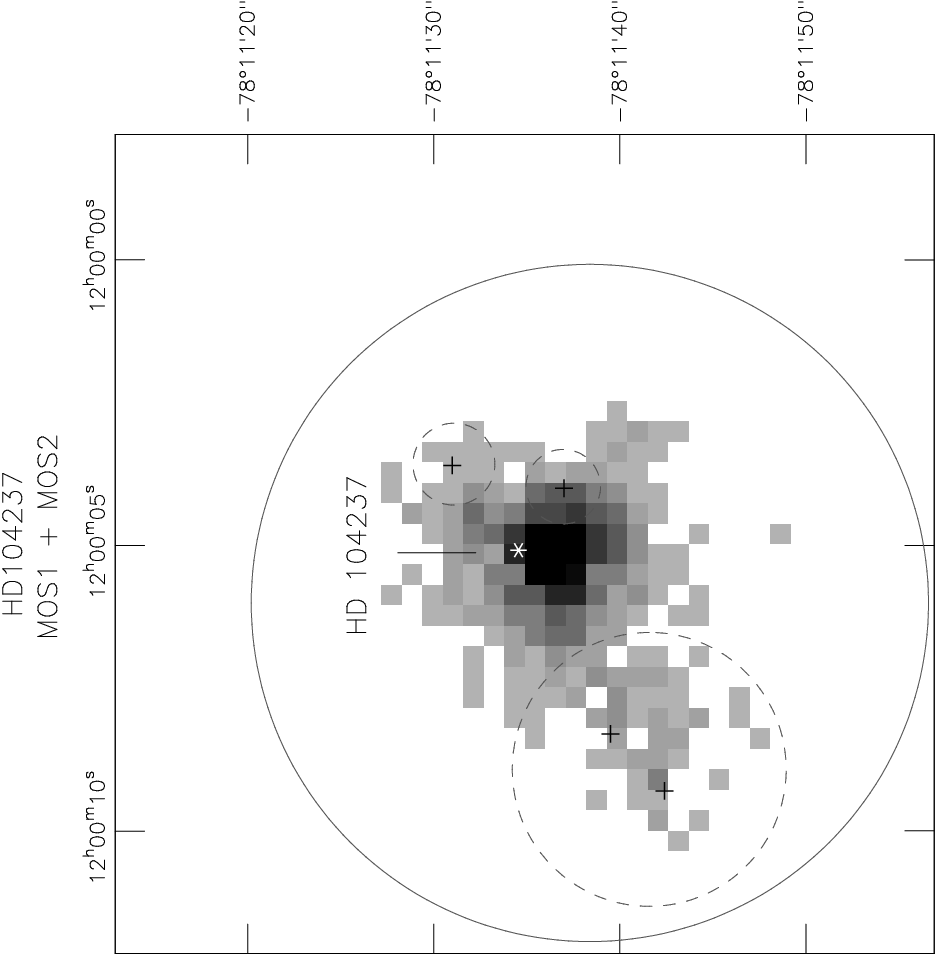}
\caption{}
\end{figure}
 
\clearpage

\begin{figure}
\figurenum{2}
\epsscale{1.0}
\plotone{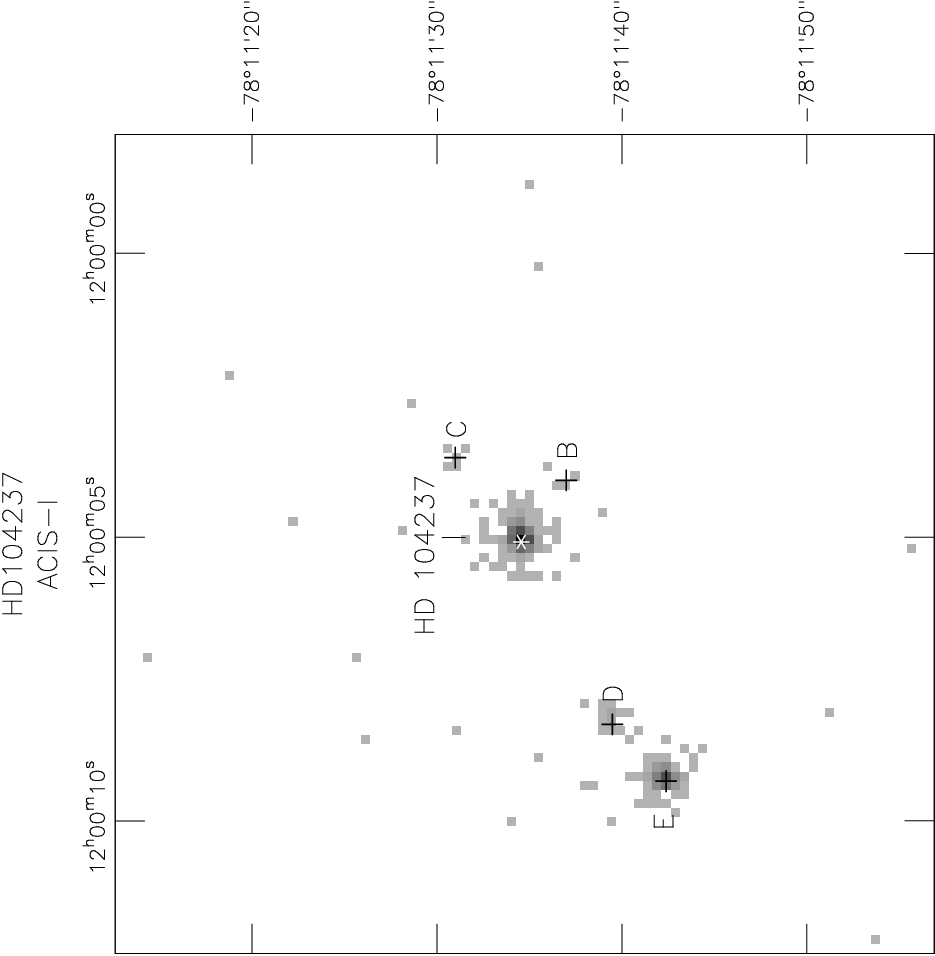}
\caption{}
\end{figure}

\clearpage

\begin{figure}
\figurenum{3}
\epsscale{0.9}
\plotone{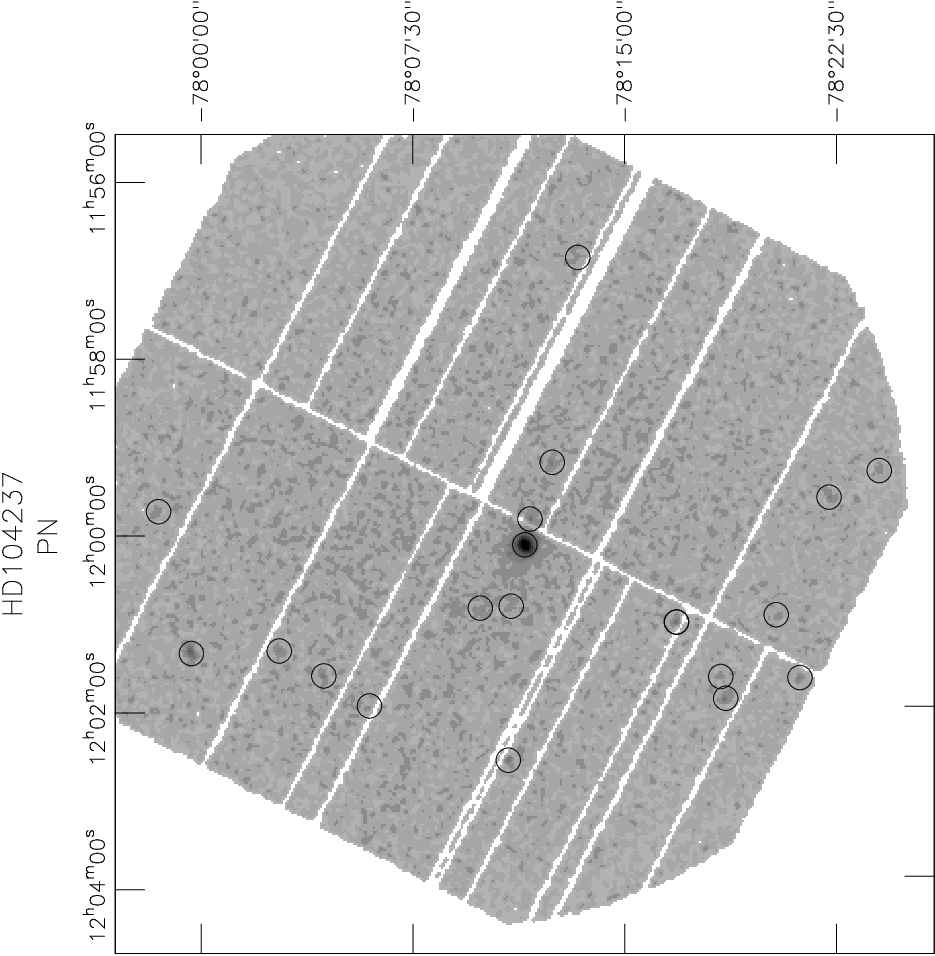}
\caption{}
\end{figure}

\clearpage

\begin{figure}
\figurenum{4}
\epsscale{0.9}
\plotone{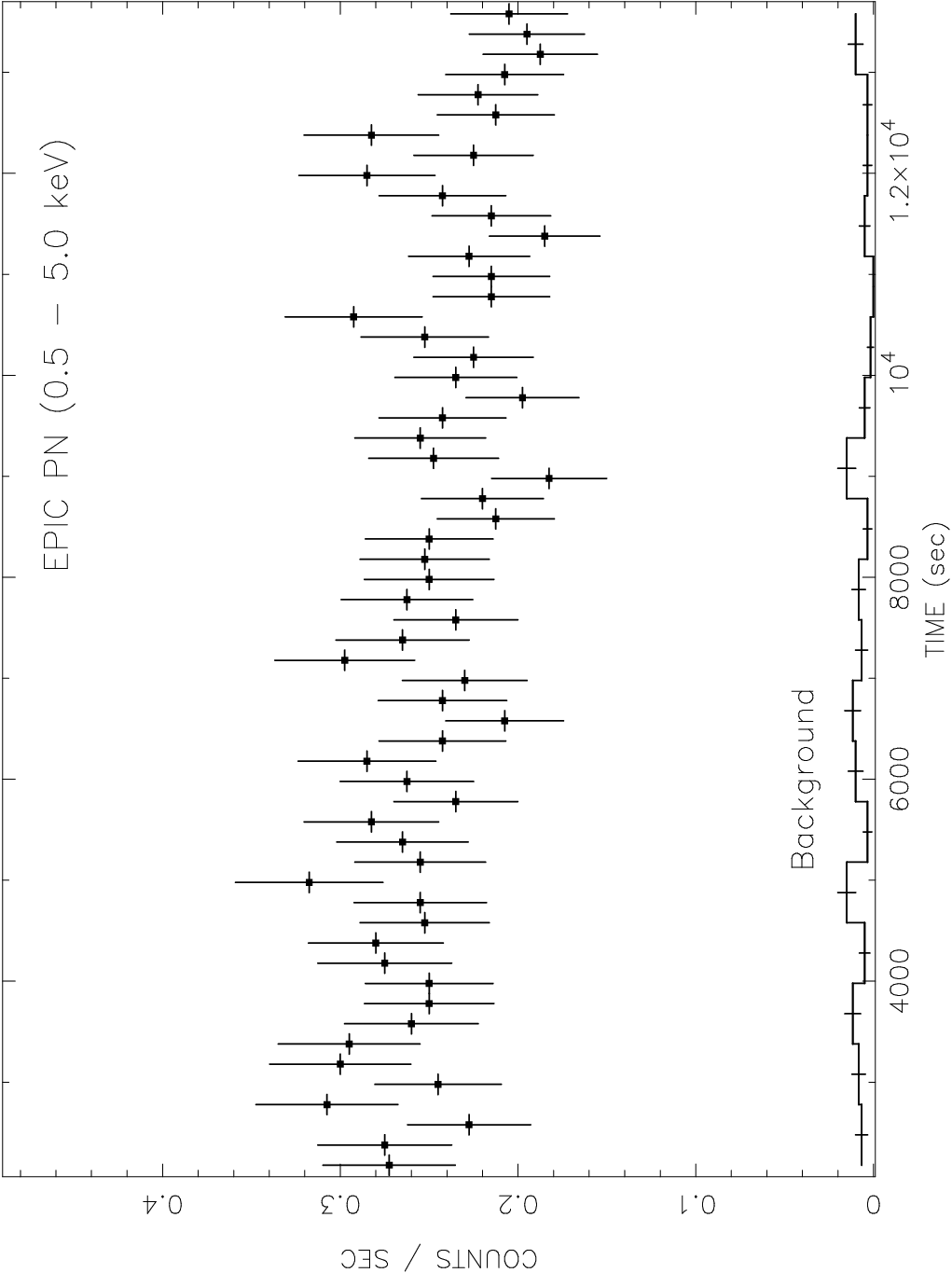}
\caption{}
\end{figure}

\clearpage

\begin{figure}
\figurenum{5}
\epsscale{0.9}
\plotone{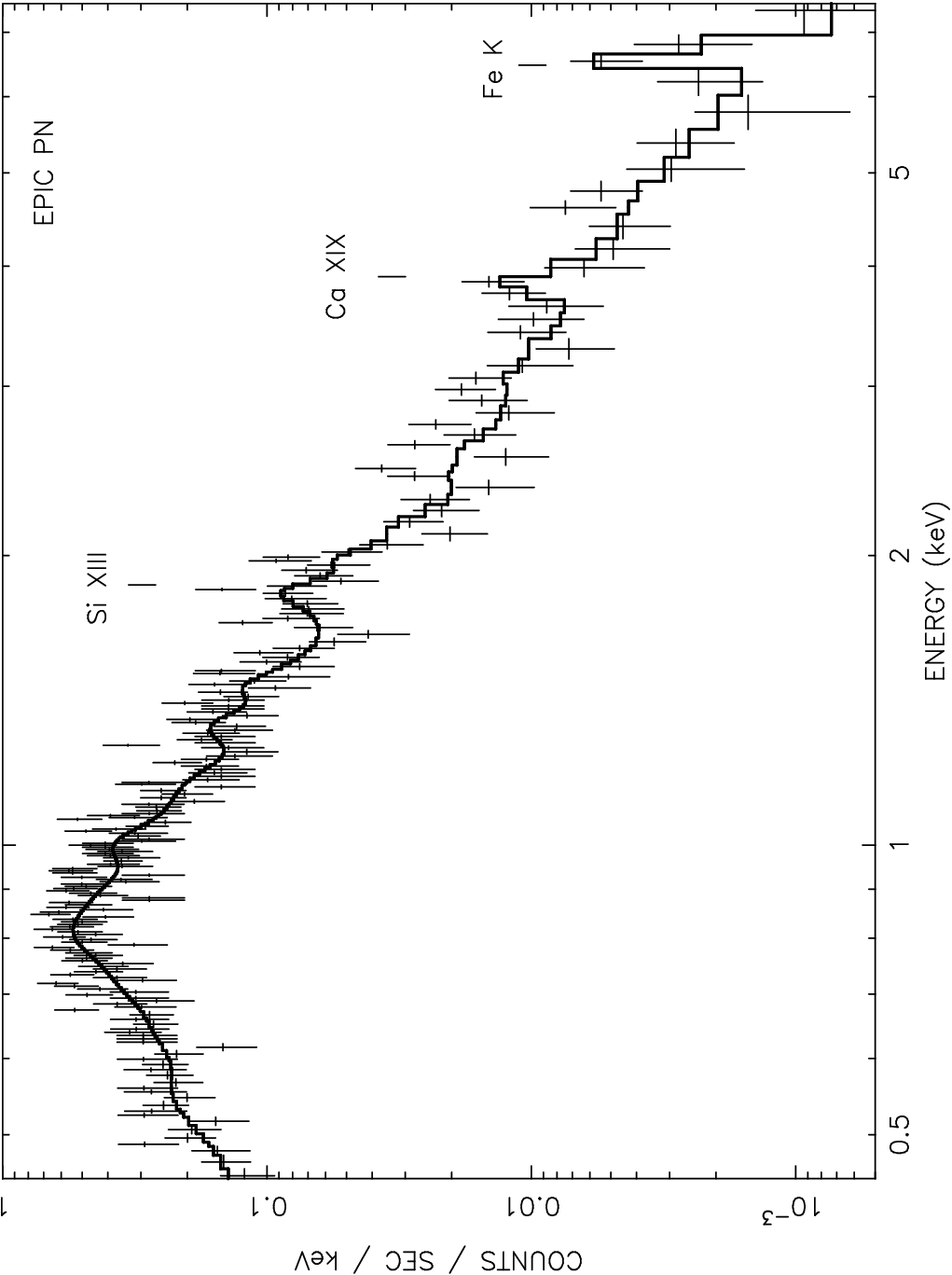}
\caption{}
\end{figure}

\clearpage

\begin{figure}
\figurenum{6}
\epsscale{1.0}
\plotone{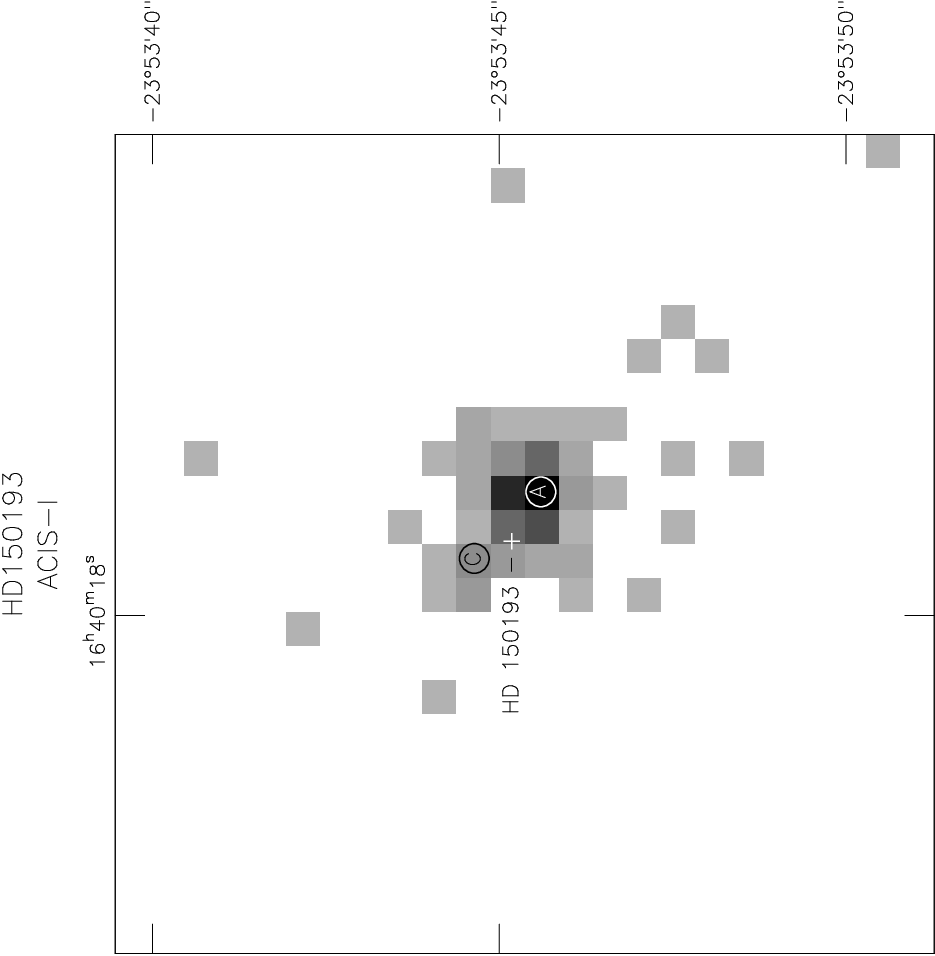}
\caption{}
\end{figure}

\clearpage

\begin{figure}
\figurenum{7}
\epsscale{0.9}
\plotone{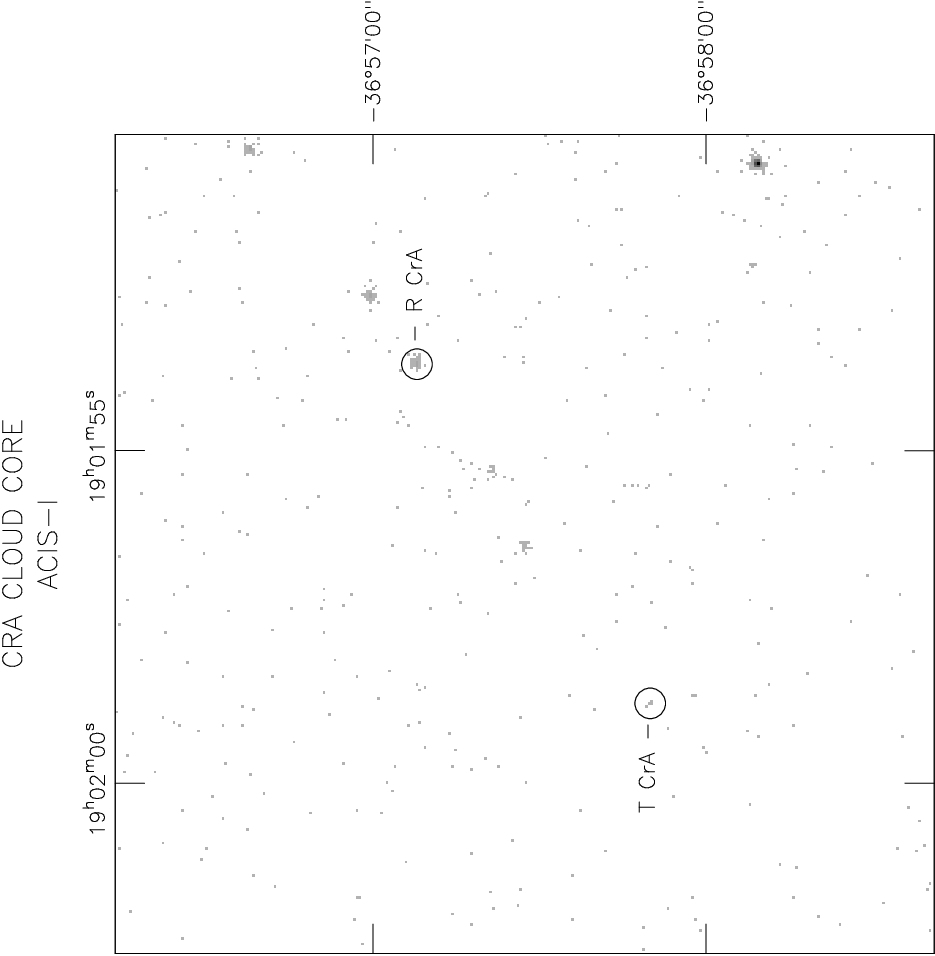}
\caption{}
\end{figure}

\clearpage

\begin{figure}
\figurenum{8}
\epsscale{1.0}
\plotone{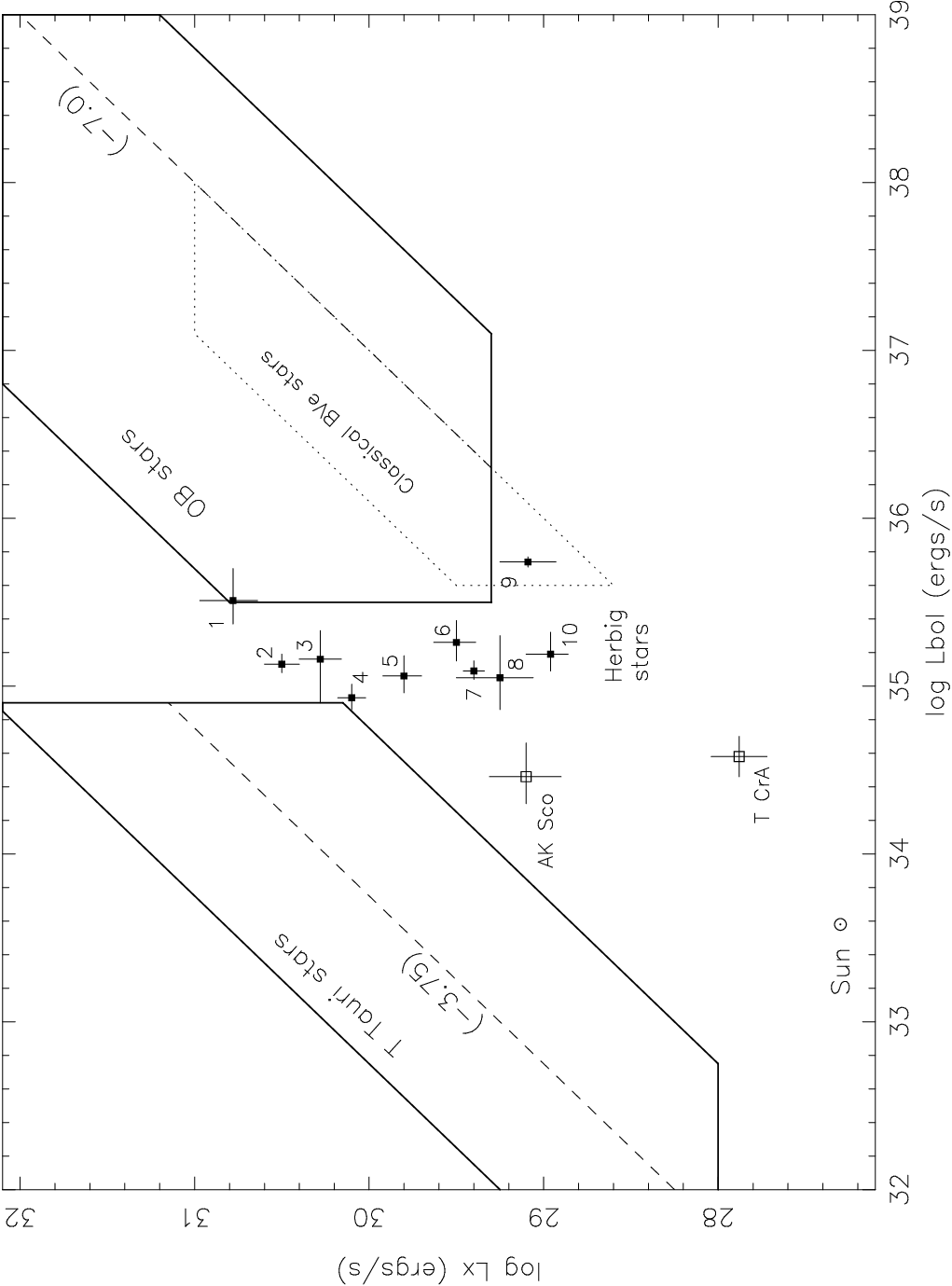}
\caption{}
\end{figure}

\end{document}